\documentclass[preprint,12pt]{elsarticle}

\usepackage{graphicx}
\usepackage{amsmath, amssymb}
\usepackage{url}
\usepackage[numbers]{natbib}
\usepackage{hyperref}
\usepackage{booktabs}
\usepackage{xcolor}
\usepackage{threeparttable}
\usepackage{tabularx}
\usepackage{longtable}
\usepackage{float}
\usepackage{colortbl}
\usepackage{array}
\usepackage{rotating}
\usepackage{multirow}
\usepackage{siunitx}
\usepackage{microtype}
\usepackage{adjustbox}
\usepackage{booktabs}

\usepackage[most]{tcolorbox}
\usepackage{soul}
\soulregister\textsc7
\soulregister\cite7

\renewcommand{\hl}[1]{#1}

\hypersetup{
    colorlinks=true,
    linkcolor=blue,
    citecolor=blue,
    urlcolor=blue
}

\journal{Computer Networks}

\begin{document}

\begin{frontmatter}

\title{Passive Reconnaissance of Routing-Layer Defenses in OLSR-Based MANETs using ML}

\author[jct]{Nadav Schweitzer}
\author[uw]{Kiril Danilchenko}
\author[jct]{Ariel Stulman}

\address[jct]{Department of Computer Science, The Jerusalem College of Technology, Jerusalem, Israel (e-mail: schwe, stulman@jct.ac.il).}
\address[uw]{Department of Computer Science, University of Waterloo, Ontario, Canada (e-mail: kirild83@gmail.com).}

\begin{abstract}
Mobile ad hoc networks (\textsc{manet}s) based on proactive routing protocols such as \textsc{olsr} remain vulnerable to routing-layer attacks. While prior work has focused primarily on attack detection, the detection of deployed defenses remains largely unexplored.

This work examines whether \hl{routing-layer defense mechanisms leave} detectable signatures in network traffic. \hl{The primary evaluation considers a fictitious-node defense against a node-isolation attack, while a second, independently proposed trust-based \textsc{fpnt} defense is evaluated under a different attack setting to examine whether passive detectability is confined to the first mechanism.}

Using ns-3 simulations across baseline, attack-only, defense-only, and combined attack-defense regimes under both static and mobile conditions, we derive features from observable routing dynamics and control-plane activity available to a passive \hl{observer confined to a single vantage point, using only frames received by its own radio interface}.

Despite \hl{this single-vantage restriction}, the results show that defense detection remains feasible in the evaluated settings. \hl{For the primary fictitious-node defense, ensemble models achieve in-domain accuracy of $0.8998$ under static conditions and $0.9298$ under mobility (\textsc{auc} $0.9403$ and $0.9729$). The independently proposed trust-based defense is likewise detectable in its respective attack setting, reaching accuracy of $0.9488$ and $0.9394$ under static and mobile conditions. Cross-domain generalization between the static and random-walk configurations remains strong in both directions: a three-observable subset yields accuracies of $0.8730$ and $0.8632$. Repeating the selection using only the source configuration recovers the same three observables in both transfer directions. A Gauss-Markov mobility control, however, produces substantially asymmetric transfer, indicating that cross-domain robustness depends on the mobility process.}

These findings indicate that the evaluated defense \hl{mechanisms can leave statistically detectable changes in observable routing behavior even from a single passive vantage point, supporting defense inference as a reconnaissance capability while limiting the evidence to the defense-attack and mobility settings evaluated here}.
\end{abstract}


\end{frontmatter}

\section{Introduction}

Mobile ad hoc networks (\textsc{manet}s) rely on decentralized routing and cooperative node behavior to maintain connectivity in dynamic topologies. Proactive routing protocols, particularly the Optimized Link State Routing (\textsc{olsr}) protocol~\cite{clausen2003rfc3626}, disseminate topology information at regular intervals, ensuring rapid route availability while exposing this control-plane information to potential manipulation by adversaries. Prior work has extensively addressed attack detection, mitigation, and classification in \textsc{olsr} and related \textsc{manet} routing protocols~\cite{hassan2024advanced}.

Despite extensive work on attack detection, relatively little attention has been given to detecting the presence of deployed defenses. In practice, during the reconnaissance stage of the cyber attack chain, attackers assess whether network conditions are favorable to the success of an attack. Defense mechanisms tailored to specific attack vectors can reduce attack effectiveness, increase the likelihood of exposing malicious activity, or neutralize the attack entirely~\cite{afraji2025deep, czaja2025cybersecurity, he2023adversarial, tebbaa2026mitigating}. \hl{Inferring whether such a defense is deployed is therefore of interest to an adversary at that stage.}

Unlike many defensive mechanisms that introduce explicit protocol modifications, the evaluated approach remains embedded within standard \textsc{olsr} control exchanges, providing no explicit signaling cues that would directly reveal its presence to a passive observer.

This distinction matters because launching an attack is a costly and partially irreversible decision. An active defense reduces the expected payoff of the attack, while exposure risk may increase due to the defense's effect on observable network behavior. A rational attacker may therefore prefer to defer or abandon the attack under such conditions. \hl{Passive reconnaissance that reveals the presence of a defense can support this decision process without any action that would expose the adversary. Whether an adversary who possesses this knowledge also achieves a better attack outcome is a separate question: it would require an adaptive adversary, which this work does not evaluate.}

To study this reconnaissance capability, we consider a passive attacker whose view is limited to information exposed through network communications. The attacker neither compromises participating nodes nor accesses internal routing information.

This work examines whether \hl{routing-layer defense mechanisms leave} detectable effects in externally observable network traffic that can be exploited by a passive adversary. To evaluate this question, we employ a machine-learning framework to analyze \textsc{olsr}-based networks across four operational regimes: (1) baseline benign operation, (2) attack-only conditions, (3) defense-only operation, and (4) combined attack-defense; each evaluated under both static and mobile network conditions. \hl{That design is applied to a fictitious-node defense against a node-isolation attack; a second, independently proposed defense is evaluated separately, under a different attack, to test whether detectability is confined to the first mechanism. Statistical descriptors derived from the traffic reaching a single passive observer are used to infer the presence of the defense, and the robustness of this inference is evaluated across the mobility settings considered here.}

\section{Related Work}
\label{subsec:Related-Work}

Prior work on \textsc{manet} security primarily focuses on detecting malicious behavior and mitigating classical routing-layer attacks, including black hole, wormhole, Sybil, and link-spoofing attacks \cite{hassan2024advanced}. Many approaches leverage anomaly-detection and machine-learning techniques to uncover inconsistencies in routing information or deviations from expected protocol behavior \cite{hassan2024advanced, saminathan2025multicast}, often in combination with graph-based representations for intrusion detection \cite{bilot2023graph}, and, in some cases, cross-layer analysis \cite{jeniffer2025efficient}. Within \textsc{manet} routing protocols, the literature also explores authentication and secure routing mechanisms to enhance resilience against malicious behavior \cite{tu2021active}. \hl{Related wireless-network research has also examined attack detection and countermeasures at lower layers, including jamming detection and mitigation in \textsc{wsn}s~}\cite{mpitziopoulos2009jamming}\hl{.}

Deploying security mechanisms such as intrusion detection and cryptographic protocols introduces additional computational costs \cite{zohourian2024iot} and communication overhead \cite{shen2024effective}, which can increase response time (latency) \cite{shen2024effective} and energy consumption \cite{sefati2025comprehensive}, impose trade-offs in bandwidth and system resources \cite{wang2025survey}, and challenge overall \textsc{q}o\textsc{s} performance in resource-constrained \textsc{i}o\textsc{t} environments \cite{sefati2025comprehensive}. Prior studies in \textsc{i}o\textsc{t} and wireless environments indicate that such defensive mechanisms result in measurable latency \cite{shen2024effective} and energy costs \cite{sefati2025comprehensive}, primarily due to additional computational \cite{zohourian2024iot} and communication requirements \cite{shen2024effective}, thereby creating trade-offs between security strength and performance efficiency \cite{sefati2025comprehensive}. While these trade-offs are typically examined from a quality-of-service or resource-constraint perspective, the resulting perturbations in network performance may also provide indirect indications of the presence of defensive mechanisms.

\hl{Observable measurements can also carry security-relevant information directly; Saedi \emph{et al.}~}\cite{saedi2022synthetic}\hl{, for example, examine measurement-report signal characteristics as potential fingerprints of rogue base stations.

Other network-integrated defenses modify protocol behavior more directly. For example, Saedi \emph{et al.}~}\cite{saedi2026zerotrust} \hl{extend the 5G handover procedure by introducing a probation period for newly observed base stations and excluding those identified as suspicious from handover decisions. While such work evaluates how reliably the defense identifies rogue base stations, these protocol-level modifications raise the converse question of whether the operation of the defense itself leaves an externally observable footprint.}

Separate lines of work examine adversarial evasion as well as the design of model-centric defenses, including adversarial training and \textsc{gan}-based defense mechanisms \cite{wang2023adversarial}. Additional studies examine evasion-aware intrusion detection and taxonomies of adversarial bypass techniques, including packet manipulation, traffic obfuscation, and payload mutation \cite{debicha2023tad, cheng2011evasion}. Reconnaissance research has investigated target topology inference \cite{hou2021combating, kong2022combination} and traffic characterization \cite{shen2022machine}.

Adversarial evasion research commonly models attackers as adaptive entities that modify their behavior in response to deployed defenses. He \emph{et al.}~\cite{he2023adversarial}, for example, formulate evasion attacks against network intrusion detection systems as test-time perturbations designed to mislead the detection model while preserving malicious functionality. Similarly, \textsc{tantra}~\cite{sharon2022tantra} considers adversaries that reshape network traffic timing patterns to match the feature space learned by machine-learning-based \textsc{ids}. While these studies demonstrate that defense-aware adaptation can substantially affect detection performance, they generally assume that the attacker already possesses knowledge of the deployed defense and do not investigate how such knowledge may be obtained during reconnaissance.

Existing studies often focus on how defensive mechanisms can be concealed or strategically deployed against adversarial reconnaissance \cite{zhu2021game, zhang2021three, pawlick2019game, kim2022equalnet, li2025tunnel}. In fact, existing studies typically assume that defensive mechanisms are either known, observable, or under the control of the defender, focusing on their design, deployment, or evasion
rather than their detectability \cite{wang2023adversarial, zhu2021game, pawlick2019game, kim2022equalnet, li2025tunnel}.

Transferring such techniques to \textsc{manet}s is challenging due to the absence of centralized control \cite{hassan2024advanced}. Without fixed infrastructure, network operation is inherently distributed \cite{singh2022cryptographic}, and nodes rely on locally available neighbor information \cite{tabatabaei2023introducing}. Consequently, unlike static networks where traffic is often routed through centralized enforcement points (e.g., network choke-points) to apply security policies \cite{kim2023extended}, \textsc{manet} defenses operate in a distributed manner and are applied locally at the node level \cite{tabatabaei2023introducing}. Furthermore, such defenses may be transient and embedded within the routing logic itself \cite{schweitzer2025achieving}, reflecting fundamental architectural differences that limit the direct applicability of security mechanisms designed for the fixed infrastructure counterparts.

To the best of our knowledge, prior work has not systematically examined whether adversaries or machine learning models can detect the presence of defenses in \textsc{olsr}-based networks, specifically under active attack scenarios, leaving this problem largely unexplored.

\section{Research Methodology}

\subsection{Defense Mechanism Overview}
\label{subsec:defense-mechanism-overview}

The present work evaluates the detectability of a defensive mechanism proposed to mitigate node isolation attacks in \textsc{olsr}-based \textsc{manet}s~\cite{schweitzer2015mitigating, schweitzer2023persuasive}. In such attacks, an adversary manipulates Multi-Point Relay (\textsc{mpr}) selection in order to distort routing topology information and disrupt node connectivity~\cite{clausen2003rfc3626}.

The evaluated defense adopts the formulation proposed by Schweitzer et~al.~\cite{schweitzer2025achieving}, in which nodes monitor local topological consistency within their 2-hop neighborhood. \hl{When this monitoring identifies a structural inconsistency,} a node injects \textit{fictitious} nodes into the perceived routing topology. This modification affects both \textsc{hello} messages, which influence local \textsc{mpr} selection, and \textsc{tc} messages, which disseminate topology information throughout the network.

This framework comprises two complementary variants: \textsc{gcop} (Graph-Coloring-for-\textsc{olsr}-Protection), which targets distance-based topological inconsistencies, and \textsc{gcohp} (Hexagon-Protection), which addresses topology-specific blind spots such as hexagonal ring structures. The two variants differ in \hl{the structural condition they evaluate} but share a common mitigation principle: controlled topology augmentation through fictitious-node injection. \hl{The defense is prophylactic: its operation follows a structural condition over the local topology rather than attack detection. It operates identically whether or not an attack is present. The structural condition is re-evaluated for each \textsc{hello} and \textsc{tc} message emitted by a node, so the set of nodes actively advertising fictitious information may change within a measurement window.}

\hl{The research question came first: whether the operation of an active \textsc{manet} defense can itself be detected through passive observation. \textsc{gcop}/\textsc{gcohp} was the first case we tried, because we had previously worked on that defense and on the node-isolation attack it counters, rather than because it was intended as a representative sample of \textsc{manet} defenses.}

\hl{To examine whether passive detectability is confined to this mechanism, we additionally evaluate \textsc{fpnt}~}\cite{tan2015trust}\hl{, an independently proposed trust-based \textsc{olsr} defense, in a black-hole attack setting. Unlike \textsc{gcop}/\textsc{gcohp}, \textsc{fpnt} does not introduce fictitious topology state; it attaches trust-related metadata concerning real neighbors to \textsc{tc} messages. The two defenses are evaluated in their respective attack settings rather than as a head-to-head comparison.}

\subsection{Classification Task Formulation}
\label{subsec:classification-task-formulation}

We formulate defense detection as a binary classification task based on externally observable routing behavior. \hl{For the \textsc{gcop}/\textsc{gcohp} evaluation, the fictive mitigation strategy}~\cite{schweitzer2015mitigating}\hl{ introduces an additive control-plane footprint by injecting fictitious nodes into the perceived topology, whereas routing attacks such as node isolation typically have degradative or subtractive effects, including packet loss, route suppression, or selective link withholding.} The dataset spans four operational scenarios: baseline, attack-only, defense-only, and combined attack-defense, each evaluated under both static and mobile network conditions.

Scenarios without an active defense (baseline and attack-only) serve as negative examples, while scenarios with an active defense (defense-only and attack-defense) represent positive instances. \hl{Here, an active defense denotes a scenario in which the defense mechanism is enabled and may operate according to its local structural condition; it does not imply that every node advertises fictitious information throughout the measurement window. The same binary formulation is applied to the \textsc{fpnt} evaluation, with its own attack setting substituted for node isolation.}

\hl{Supervised training uses labelled data from the source configuration only; the target configuration is evaluated without labels. The balanced class construction provides a controlled test of whether the defense states are statistically separable; it does not represent their prevalence in an operational network. Accordingly, accuracy under this balanced design measures separability rather than an operational detection rate.}

\hl{The adversarial model assumes a passive observer that adds no traffic or interaction beyond its normal protocol participation. Features are restricted to control-plane and data-plane frames received by the observer's own radio interface and to quantities computable solely from those frames. No cross-node aggregation or reconstruction of network-wide quantities is assumed.} Metrics requiring access to node-internal state or non-observable measurements (e.g., routing tables, MAC-layer internal counters, hardware-dependent quantities, or physical mobility measurements) are excluded.

\hl{The threat model also fixes the defender's behaviour. The evaluated defense acts only on its local structural condition and does not alter what it emits in order to conceal its own footprint from a passive observer; no concealment-aware variant of the defense is evaluated here.}

\hl{The observer has only a single spatial vantage point and therefore does not have complete network visibility. Although \textsc{olsr} flooding may allow it to receive \textsc{tc} messages originating beyond its immediate neighborhood, we do not assume that every such message reaches the observer. The restriction is spatial: the observer uses only the frames that reach its own position. Within that region the propagation model is range-based, so distance-dependent attenuation and fading are not represented; medium-access contention and collisions, including those arising from hidden terminals, are modeled as in standard IEEE 802.11.}

\subsection{Metric Collection and Feature Engineering}
\label{subsec:metric-collection}

All metrics were derived from packet-level traces collected from ns-3~\cite{ns3} simulations, \hl{under the single-vantage observation model defined in Section~}\ref{subsec:classification-task-formulation}\hl{.}

\hl{The simulator records a broad set of per-window network metrics commonly used to characterize \textsc{manet} performance, including routing overhead, packet delivery ratio, throughput, delay, jitter, and energy consumption}~\cite{quy2019survey,eltahlawy2023survey}\hl{.} Consistent with the adversarial model defined in Section~\ref{subsec:classification-task-formulation}, only metrics derivable from \hl{frames received by the observer's own radio interface} were retained. Consequently, hardware-dependent quantities, \textsc{mac}-layer internal counters, node-internal routing state, physical mobility measurements, \hl{and quantities requiring end-to-end or network-wide information, such as delay, jitter, and delivery and loss quantities,} were excluded. \hl{Quantities whose only single-vantage analogue depends materially on the observer's reception range were also excluded because they do not faithfully represent their network-wide counterparts; retained control-plane quantities are instead derived from the advertised content of flooded \textsc{tc} messages received by the observer.}

\hl{This filtering yields 17 single-vantage observables derived from control-plane and data-plane traffic received by the observer, which form the common basis for all downstream analyses.}

\hl{Feature expansion was guided by domain knowledge: interactions and ratios represent control-plane amplification and protocol-level imbalances; logarithmic, square-root, and polynomial transformations represent non-linear scaling; and row-wise summary statistics capture joint variation within each measurement window. The resulting expansion contains 67 columns: 17 original observables, 17 logarithmic transformations, 17 square-root transformations, two products and sums of the observables, twelve row-wise summary statistics, and two polynomial terms. The feature-engineering code, released with the processed data at the repository given in Section~}\ref{subsec:evaluation_reproducibility}\hl{, provides the complete specification of the generated columns together with the retained-column lists for both configurations.}

\hl{Feature selection first discarded attributes with variance below $0.01$, then removed one feature from each pair with Pearson $|r|>0.95$ to reduce near-duplicate information}~\cite{solorio2022survey}\hl{. These filtering steps were fitted on the training partition.}

After filtering, the remaining engineered features were ranked by aggregating four complementary importance criteria: mutual information with the class label, \textsc{anova} F-statistic, Random Forest feature importance, and ExtraTrees feature importance. The final rank for each feature was computed as the unweighted mean of its four criterion-specific ranks, and the top-ranked features were retained for classifier training. \hl{This process retains 28 features for the static configuration and 32 for the mobile configuration.}

Unless stated otherwise, the extracted features represent aggregated observations over time rather than instantaneous measurements.

\subsection{Feature Taxonomy}
\label{subsec:feature_taxonomy}

The retained observable metrics, together with their derived feature representations, are grouped into \hl{three categories according to the network processes they capture: control-plane behavior, locally observed traffic activity, and per-source traffic dynamics. The taxonomy is defined at the level of these underlying observable processes rather than individual engineered variables.}

\subsubsection{\hl{Control-Plane Topology and Overhead}}

Since Fictive Mitigation introduces fictitious topology information into the routing process, its effects are expected to appear primarily in control-plane behavior. These effects may influence topology advertisements \hl{and routing overhead}.

\hl{Four control-plane metrics reappear throughout the subsequent analysis and are therefore described briefly below: TcMessageRate, AverageAdvertisedLinksPerTCMessage, AdvertisedLinksPerKnownNode, and AverageHopCount.}

\paragraph{\hl{TcMessageRate}}

\hl{TcMessageRate measures the rate of distinct topology control (\textsc{tc}) messages received by the observer during the measurement window. Flooded copies are de-duplicated by the pair (originator, message sequence number) for this metric. It therefore captures changes in the volume of topology information visible from the observer's single vantage point.}

\paragraph{AverageAdvertisedLinksPerTCMessage}

AverageAdvertisedLinksPerTCMessage measures the average number of links advertised in topology control (\textsc{tc}) messages \hl{received by the observer}. \hl{Specifically, if $R$ is the total number of advertised links carried by all received \textsc{tc} messages, with flooded copies counted separately, and $M$ is the corresponding number of received \textsc{tc} messages, the metric is computed as $R/M$.}

\paragraph{\hl{AdvertisedLinksPerKnownNode}}

\hl{AdvertisedLinksPerKnownNode measures the advertised-link state most recently observed from the \textsc{tc} originators relative to the number of distinct node addresses seen by the observer. For each originator $i$, let $a_i$ denote the number of links carried in its most recently received \textsc{tc} message, let \mbox{$S=\sum_i a_i$}, and let $D$ be the number of distinct node addresses appearing either as a \textsc{tc} originator or within an advertised link set. The metric is computed as $S/D$.}

\hl{Because \textsc{tc} messages disseminate topology-state information beyond the originator's immediate neighborhood, changes in the perceived topology may affect the amount of information carried in each advertisement. AverageAdvertisedLinksPerTCMessage therefore serves as an indicator of changes in the advertised network topology.}

\paragraph{\hl{AverageHopCount}}

\hl{AverageHopCount is the mean value of the hop-count field carried in the topology control (\textsc{tc}) messages the observer receives. It is a property of the control messages heard at the observer and not the number of hops taken by data traffic.}

\subsubsection{\hl{Locally Observed Traffic Characteristics}}

\hl{The retained feature space also includes metrics describing the frames and data traffic received by the observer, such as their rate and size.}

\subsubsection{\hl{Per-Source Traffic Statistics}}

\hl{Per-source metrics characterize the activity and variability of source addresses visible to the observer, such as variation in traffic volume across sources. They do not require reconstruction of end-to-end flows and are computed solely from locally received traffic.}

\hl{To limit dependence on defense-specific feature choices, the analysis applies the single-vantage observability audit and the documented feature-generation procedure described above, together with the sensitivity analysis across importance criteria reported in Section~}\ref{subsec:feature-consistency}\hl{.}

\section{Simulation Environment and Network Configuration}
\label{sec:simulation-environment}

The experimental evaluation was conducted using the \hl{ns-3.47} network simulator~\cite{ns3}. The network included $N = 50$ nodes whose initial positions were placed at random over a $750 \times 1000$\,m$^2$ area, operating under the \textsc{olsr} routing protocol with the fictive mitigation mechanism \hl{implemented}.
Node roles were fixed across runs, with a designated victim node and a consistent traffic source. The attacker was deterministically selected among the victim’s one-hop neighbors at the time of activation.
These parameters were selected to ensure a stable multi-hop regime while maintaining sufficient routing dynamics to expose control-plane effects induced by both attack and defense mechanisms.
\hl{For each run, the observer is one of the nodes holding no fixed role in the scenario, selected uniformly at random.}

In the mobile configuration, nodes follow a \texttt{RandomWalk2dMobilityModel} with movement speeds drawn randomly from $[1.5, 2.0]$ m/s ($5.4-7.2~km/h$), with direction changes applied every 3\,s and reflective boundaries within the deployment area. For the static configuration, mobility is disabled (\texttt{ConstantPositionMobilityModel}) to isolate protocol behavior from topology dynamics. The initial node positions are randomly placed on the deployment area in both configurations. All mobility-related random streams are deterministically derived from the per-run \texttt{RngRun} seed. Full mobility parameters and simulation setup are summarized in Table~\ref{tab:simulation-parameters}.

\begin{table}[h]
\centering
\caption{Summary of Simulation Parameters}
\label{tab:simulation-parameters}
\resizebox{\columnwidth}{!}{%
\begin{tabular}{ll}
\hline
Category & Value \\
\hline
Simulator                           & \hl{ns-3.47} \\
Routing protocol                    & \textsc{olsr} \\
Defense mechanism                   & Fictive Mitigation \\
Nodes                               & 50 \\
Deployment area                     & 750 $\times$ 1000 m\textsuperscript{2} \\
Wireless standard                   & IEEE 802.11a (ad-hoc) \\
PHY rate                            & 6 Mbps (fixed) \\
Propagation model                   & \hl{Range-based (main campaign); log-distance + Nakagami (sensitivity analysis)} \\
Transmission range                  & 190 m \\
Traffic type                        & UDP constant bit-rate \\
Measurement window duration         & 40s \\
Simulation duration                 & 400s \\
Simulation runs (per configuration) & 10,000 \\
Total simulation runs               & 20,000 \\
Total samples                       & 80,000 \\
Mobility model                      & \texttt{RandomWalk2dMobilityModel} \hl{(main campaign); \texttt{Gauss-Markov}, \mbox{$\alpha = 0.85$} (mobility control)} \\
Initial positions                   & Uniform random \\
Node speed                          & Uniform [1.5, 2.0] m/s \\
\hl{Mobility update interval}       & 3s \hl{(both mobile configurations)} \\
Boundary handling                   & Reflection at edges \\
\hline
\end{tabular}%
}
\end{table}

\hl{A second mobility model is evaluated as a control. A further campaign of $10{,}000$ accepted runs replaces \texttt{RandomWalk2dMobilityModel} with the Gauss-Markov mobility model, leaving the node count, deployment area, wireless configuration, traffic pattern, node-speed distribution, single-vantage observation model, admission criterion, feature space and pipeline of the main campaign unchanged. Its static half is the static campaign reported here, so the two campaigns differ only in the mobility process of the mobile configuration. The Gauss-Markov arm uses a time step of $3$\,s, matching the direction-update interval of the random-walk configuration, and a memory parameter \mbox{$\alpha = 0.85$}, corresponding to a characteristic memory-decay time of approximately $18$\,s. Its mean velocity is drawn from the same \mbox{$[1.5, 2.0]$}\,m/s distribution as the random-walk configuration, so the two mobile arms differ in the temporal structure of the motion and not in node speed. Mean direction is uniform on \mbox{$[0, 2\pi)$}, the velocity and direction increments are zero-mean normal with variances $0.04$ and $0.2$ and bounds $0.4$ and $0.8$, respectively, and pitch is fixed at zero, keeping the scenario two-dimensional. The value of \mbox{$\alpha$} is a declared modelling choice rather than a fitted value.}

Wireless communication followed the IEEE 802.11a ad-hoc model with a fixed 6\,Mbps data rate and \hl{a range-based propagation-loss model with} an effective transmission range of approximately $190$\,m, resulting in a multi-hop topology. \hl{Medium-access contention and collisions are modeled throughout; RTS/CTS was not enabled, so hidden-terminal collisions are present in all reported campaigns, and their effect is examined separately in Section~}\ref{subsec:robustness}\hl{.} A \textsc{udp} constant bit-rate traffic flow was established toward the victim node, transmitting 512-byte packets every 2\,s. \hl{This is the data traffic that the Node Isolation Attack denies; the packet count and packet size follow the measurement convention of the works in which the evaluated defense was introduced, where the fraction of the eighteen packets reaching the destination is reported as the packet delivery ratio~}\cite{schweitzer2023persuasive, schweitzer2025achieving}\hl{. The measurement window here is 40\,s rather than the 300\,s used in those works.} \hl{For this flow, only quantities computed from frames received by the observer enter the feature space}, while most features were derived from control-plane activity and routing dynamics\hl{; data-plane observables are retained in the feature space, while the dominant discriminative signal resides in the selected control-plane features}.

To evaluate adversarial conditions, a targeted Node Isolation Attack was activated. An attacker from the victim's immediate neighbors forged neighbor advertisements and omitted the victim from topology control messages, thereby removing it from the routing graph.

\hl{A second evaluation used \textsc{fpnt} under its corresponding black-hole attack setting~}\cite{tan2015trust}\hl{. The same node count, deployment area, wireless configuration, traffic pattern, single-vantage observation model, and the same full-connectivity admission criterion as the main campaign were used, with no minimum source-victim distance filter. The black-hole attacker was selected as an on-path forwarder between the traffic source and the victim. The \textsc{fpnt} campaign comprised 2,000 accepted runs in each of the static and mobile modes, producing 8,000 measurement windows per mode. Its run seeds were a declared subset of the corresponding campaign population, taken in manifest order. Feature selection, model fitting, probability calibration, and decision-threshold selection were performed independently on the \textsc{fpnt} data rather than transferred from the \textsc{gcop}/\textsc{gcohp} evaluation. In the \textsc{fpnt} campaign, the designated node 2 was placed near the centre of the deployment area and, under mobility, kept stationary there. Because \textsc{fpnt} and \textsc{gcop}/\textsc{gcohp} were evaluated against different attacks, their results characterize the two defense-attack settings separately rather than constituting a controlled comparison between the defenses.}

\subsection{Measurement Procedure}
\label{subsec:measurement-procedure}

\hl{Each simulation in the main campaign} lasted 400\,s. The first 60\,s were reserved for initial routing convergence. Convergence was verified by confirming that each node maintained complete routing information to all other nodes.

Four, non-overlapping, 40\,s measurement windows were recorded within each run, corresponding to: (1) baseline operation, (2) attack-only conditions, (3) defense-only operation, and (4) combined attack-defense conditions. A 60\,s stabilization period was introduced between each of the windows, to allow the network to settle after each configuration change. All counters were reset at the beginning of each window \hl{so that each sample contained only observations recorded during that window}. Each window was treated as a labeled sample. \hl{The main campaign used this fixed order, thereby confounding scenario identity with elapsed simulation time. As a control, the four conditions were randomly permuted per seed across the four measurement positions, with the fixed-order and permuted arms using the same accepted seeds. Both arms used a 60\,s formation lead-in and retained the 60\,s stabilization period at every position, making their runs 60\,s longer than those of the main campaign. Detection accuracy is close in the two arms: $0.8788$ for the fixed order against $0.8831$ for the permuted arm under static conditions, and $0.9038$ against $0.9000$ under mobility. Both arms are scored with the stacking ensemble on the same $400$ test runs, and the paired difference, computed as permuted minus fixed and resampled at the simulation-run level, is $+0.0044$ with a $95\%$ interval of \mbox{$[-0.0050, +0.0144]$} under static conditions and $-0.0038$ with \mbox{$[-0.0175, +0.0106]$} under mobility. Each interval contains zero and has a half width of about one to $1.4$ accuracy points, so the comparison does not resolve a difference at this precision rather than establishing that none exists.}

Within each measurement window, the \textsc{udp} flow was configured to transmit 18 packets. \hl{Traffic generation was initiated 4\,s after the beginning of each window to avoid transient effects; the eighteen datagrams therefore span the remaining 36\,s at the 2\,s interval. To test whether this packet count limits what is observable, the evaluation was repeated at two higher traffic densities, varying only the inter-packet interval, over paired $2{,}000$-run campaigns sharing one seed set. At $2.0$\,s, $1.0$\,s and $0.5$\,s, corresponding to $18$, $36$ and $72$ datagrams per window, accuracy is $0.8856$, $0.8913$ and $0.8894$ under static conditions and $0.9063$, $0.9000$ and $0.9063$ under mobility. Across this fourfold increase, accuracy varies by only $0.0056$ and $0.0063$ under static and mobile conditions, respectively. The sweep varies constant-bit-rate packet density only; it does not test bursty or heterogeneous traffic.}

\hl{The \mbox{$40$\,s} window is the evaluated operating point rather than a minimum observation requirement, and its length was varied to establish that. Windows of $10$, $20$ and $40$\,s were evaluated in separate $2{,}000$-run campaigns, since the window is a simulation-time parameter and cannot be obtained by slicing a longer one after the fact. Measured against the $40$\,s condition by resampling the simulation runs the campaigns share, the difference is \mbox{$+0.38$} percentage points \mbox{$[-0.75, +1.56]$} at $20$\,s and \mbox{$+0.94$} \mbox{$[-0.13, +1.94]$} at $10$\,s under static conditions, and \mbox{$-0.75$} \mbox{$[-2.50, +0.88]$} and \mbox{$-2.56$} \mbox{$[-4.50, -0.63]$} under mobility. Only the $10$\,s mobile interval excludes zero. The other three leave room for a change of one to two accuracy points in either direction and are not read as equivalence, but no cost is resolved at $20$\,s in either configuration, so a shorter observation budget than the one evaluated here still carries the signal. The sensitivity to the size of the training corpus was also evaluated. Holding the reported estimator fixed, a $2{,}000$-run subsample of the same campaign reduces the training partition from $6{,}000$ to $1{,}200$ runs and lowers measured accuracy by $1.42$ percentage points under static conditions and $2.35$ under mobility.}

\hl{A further control varies how much of the measurement window contains defense activity. The defense is permitted to operate normally for the first fraction $f \in \{0.25, 0.50, 0.75, 1.00\}$ of each 40\,s window and is disabled thereafter. Within that portion it still applies its structural condition to each emitted message, so $f$ bounds the time the defense may act rather than the proportion of messages carrying fictitious information. The detector trained under ordinary operation is held fixed, without refitting, reselection, recalibration, or rethresholding, and all conditions use the same test-partition seeds ($411$ static, $396$ mobile). Under mobility, accuracy decreases from $0.9293$ at $f=1$ to $0.8832$ at $f=0.25$, a paired difference of $-0.0461$ $[-0.0587,-0.0354]$. Static results show no consistent trend: the differences at $f=0.25$ and $f=0.75$ include zero, whereas $f=0.50$ exceeds continuous operation by $+0.0219$ $[+0.0128,+0.0316]$. Because shortening the active period changes both the amount and temporal structure of defense activity, the static result is not interpreted as evidence that shorter activation is beneficial. Detection therefore persists when the defense is active for only part of the window, over the range evaluated here.}

Simulations were conducted separately under static and mobile configurations. In each configuration, every simulation run used a single random seed, shared across all four measurement windows, \hl{so that the four scenarios within a run belonged to the same random realization}.
\hl{The windows therefore share one realization of the node placement and, under mobility, of the movement process. The topology remains fixed across scenarios in the static configuration, whereas under mobility it continues to evolve within the shared realization.}

\hl{Every run contributes examples from both classes, and partitioning is performed at the simulation-run level, so no run contributes windows to both training and test data. A further independent-window control comprised nearly 16{,}000 simulations, with independently seeded scenarios drawn without replacement from the same accepted population. Only the first measurement window of each simulation was retained, following a 60\,s formation lead-in, thereby removing both the shared-seed structure and accumulated protocol or queue state from preceding phases. Detection remains strong on that control: accuracy is $0.8888$ (\textsc{auc} $0.9277$) under static conditions and $0.9137$ (\textsc{auc} $0.9548$) under mobility. Because its windows are drawn from independent topologies and its timeline carries a formation lead-in, it is not paired with the main campaign, and no difference between the two is read as an effect size.}

For each retained feature, total variance is decomposed into a within-run component (variation across the four windows of a given run) and a between-run component (variation of per-run means). \hl{The ratio $R = \sigma^2_{\text{within}} / \sigma^2_{\text{between}}$ is retained as a descriptive measure rather than a feature-selection or leakage criterion. For the 28 retained static features, $R$ ranges from $0.0144$ to $1.7115$ (median $0.5543$), with 11 below the descriptive $0.5$ cutoff; for the 32 mobile features, it ranges from $0.2031$ to $2.5168$ (median $1.2710$), with 5 below the cutoff. Thus, the variance decomposition does not by itself establish the absence of topology-correlated variation; protection against cross-run leakage comes from the run-grouped partition, and the independent-window control tests the concern empirically.}

\subsection{Dataset Construction}
\label{subsec:dataset-construction}

The following section describes how simulation runs were filtered and assembled into the final dataset used for classification.

Simulation runs were generated using distinct random seeds until $10{,}000$ valid runs were obtained for each configuration (static and mobile). \hl{Table~}\ref{tab:attempt_stats}\hl{ summarizes the attempt counts reconstructed from the acceptance records and the corresponding acceptance rates.}

\begin{table}
\centering
\caption{Simulation Attempts and Acceptance Rates.}
\label{tab:attempt_stats}
\begin{tabular}{lrrrr}
\toprule
Configuration & Attempts & Accepted & Rejected & Accept rate \\
\midrule
Static & \hl{19{,}932} & 10{,}000 & \hl{9{,}932} & \hl{50.2\%} \\
Mobile & \hl{28{,}241} & 10{,}000 & \hl{18{,}241} & \hl{35.4\%} \\
\bottomrule
\end{tabular}
\end{table}

A run was retained in the analysis dataset \hl{if, by the end of the initial stabilization period, every node could reach every other node through \textsc{olsr}.}

\hl{Execution of the Node Isolation Attack additionally required at least one node other than the traffic source to be a one-hop neighbor of the victim, providing a candidate attacker in the position required by the attack.}

\hl{In $27.3\%$ of the admitted static runs and $33.5\%$ of the admitted mobile runs, the traffic source is within two hops of the victim in at least one measurement window.}

A potential concern is that the \hl{acceptance criterion} may introduce systematic topological bias: accepted runs could share structural properties (e.g., denser neighborhoods or shorter average path lengths) that make defense detection easier than in a truly random deployment, inflating the reported performance.

To assess whether accepted and rejected runs differ structurally, we compare them using three topology-level metrics that are not used directly in the \hl{acceptance criterion}: average neighbor count, two-hop neighborhood size, and average minimum Euclidean distance to a neighbor. These metrics capture complementary aspects of network structure, including local connectivity, two-hop neighborhood density, and geometric dispersion, providing coverage of the primary topological factors relevant to the \hl{acceptance criterion}.

\begin{table}
\centering
\caption{\hl{Cohen's $d$ effect sizes between accepted and rejected runs on three topology metrics not used directly in the acceptance criterion, computed over the seeds probed to assemble the reported dataset. Positive $d$ indicates a larger value among accepted runs. Effect-size categories follow Cohen's conventions: $|d|<0.2$ trivial, $0.2 \leq |d| <0.5$ small.}}
\label{tab:cohen_d_topology}
\begin{tabular}{lcc}
\toprule
Metric & Static $d$ & Mobile $d$ \\
\midrule
Avg.\ neighbor count                       & \hl{$-0.140$} & \hl{$-0.066$} \\
Avg.\ two-hop neighborhood size            & \hl{$-0.265$} & \hl{$-0.171$} \\
Avg.\ min.\ Euclidean dist.\ to neighbor   & \hl{$+0.075$} & \hl{$+0.126$} \\
\bottomrule
\end{tabular}
\end{table}

The observed effect sizes are small or negligible. \hl{In the mobile configuration, all three metrics exhibit negligible differences (max $|d| = 0.171$). In the static configuration, two metrics remain negligible ($|d| \leq 0.14$), whereas one (average two-hop neighborhood size) demonstrates a small effect ($d = -0.265$).} Overall, the examined topology metrics exhibit only negligible or small differences between accepted and rejected runs, suggesting that the filtering does not systematically favor topologies with substantially different connectivity characteristics.

Beyond these measurements, direct comparison of accepted and rejected runs on arbitrary structural properties is not possible, as rejected runs are not retained in the final dataset. \hl{However, even if we were to allow for residual bias on metrics not tested, acceptance is determined at the run level rather than per scenario: a given seed is either accepted for all four measurement types or rejected entirely. The positive (defense-active) and negative (defense-inactive) classes therefore draw from the same set of accepted runs. Any structural bias introduced by the filtering therefore applies to both classes, although the accepted population may still differ structurally from the excluded population.}

\hl{The reconstructed acceptance records correspond to rejection rates of $49.8\%$ and $64.6\%$ in the static and mobile configurations, respectively. The substantial rejection rate, particularly under mobility, limits the scope of the reported results. The excluded population consists of realizations in which the sampled node placement and, under mobility, the positions reached by the end of stabilization do not provide network-wide reachability. Our conclusions therefore apply to the connected regime represented by our experimental configuration, and are not claimed to extend to partitioned or intermittently connected networks.}

Each accepted run produced four labeled samples, one for each measurement window. The final dataset is therefore constructed by aggregating window-level samples across all runs, yielding a balanced representation of the four scenarios.

\subsection{Feature Extraction}
\label{subsec:feature-extraction}

Features were extracted from packet-level traces for each measurement window. \hl{The single-vantage representation comprises 17 distinct observables, each computed solely from frames received by the observer's radio; all subsequently generated features are constructed exclusively from these observables.}

In-domain classification experiments (Section~\ref{subsec:in-domain}) used an expanded \hl{67-dimensional feature space generated from these 17 observables, composed as described in Section~}\ref{subsec:metric-collection}\hl{.} Feature selection was then performed independently for each configuration, yielding \hl{28} retained features for the static configuration and \hl{32} for the mobile configuration, using only the respective training partitions. \hl{Validation data were split at the run level into disjoint subsets for probability calibration and threshold tuning, while the test partition was reserved exclusively for final evaluation.}

Cross-domain analyses in Sections~\ref{subsec:cross-domain} through~\ref{subsec:feature-ablation-and-augmentation} used the \hl{17 base observable metrics} without further expansion, and compact universal subsets were derived through the \hl{repeated-resampling} procedure described in Section~\ref{subsec:feature-consistency}.

\section{Classification Framework}
\label{sec:classification_framework}

\hl{Each measurement window is one labeled sample, under the labeling scheme of Section~}\ref{subsec:classification-task-formulation}\hl{.} Since each simulation run produces two samples from each class, the dataset is balanced by construction with a $1{:}1$ class ratio.

\subsection{Data Partitioning}
\label{subsec:data_partitioning}

Classification is performed separately for the static and mobile configurations. Each configuration contains $10{,}000$ accepted runs, corresponding to $40{,}000$ labeled samples.

The dataset is divided into training, validation, and test partitions using a $60/20/20$ split at the run level, \hl{implemented with grouped random sampling (}\texttt{GroupShuffleSplit}\hl{) keyed on the simulation run, so all four measurement windows of a run remain in the same partition}, corresponding to $24{,}000$ training, $8{,}000$ validation, and $8{,}000$ test samples per configuration. Class balance is preserved in all partitions. All numerical features are standardized using a robust scaler fitted on the training partition.

\subsection{Classifiers and Hyperparameter Configuration}
\label{subsec:classifiers-hyperparameters}

\hl{We evaluate sixteen classifier configurations spanning five model families: tree-based ensembles (Random Forest, Extra Trees, Gradient Boosting), gradient boosting libraries (two configurations each of \textsc{xgboost}, \textsc{catboost}, and \textsc{lightgbm}), bagging and boosting meta-estimators (AdaBoost, Bagging-RF, Bagging-ET), linear models (Logistic Regression, Ridge), and support vector machines (an \textsc{rbf}-kernel \textsc{svm} using \textsc{libsvm}, and a linear \textsc{svm} with squared hinge loss using a liblinear-family solver).} Tree-based and boosting methods are emphasized due to their effectiveness on heterogeneous tabular data~\cite{shwartz2022tabular,grinsztajn2022why}, while linear and kernel-based models provide reference baselines.

Two further linear classifiers, a linear \textsc{svm} with hinge loss and Linear Discriminant Analysis (\textsc{lda}), are \hl{evaluated as additional baselines} in Section~\ref{subsec:in-domain}\hl{. Both are fitted on the training partition and calibrated and thresholded under the same protocol as the models above, but are not among the sixteen classifier configurations and do not enter the Stacking Ensemble.}

\hl{Hyperparameter configurations were selected in an earlier round of this study using a unified search procedure applied independently to the static and mobile configurations.} For tree-based, boosting, and bagging models, \texttt{RandomizedSearchCV} with $n_{\text{iter}}=80$ \hl{was} applied; for Logistic Regression, Ridge, and \textsc{svm}, \texttt{GridSearchCV} \hl{was} used~\cite{probst2019tunability}. All searches \hl{employed} 3-fold stratified cross-validation on the training partition with accuracy as the scoring criterion. The procedure \hl{was} deterministic with $\texttt{random\_state}=42$.

\hl{Search spaces covering standard ranges for the number of estimators, tree depth, learning rates, and regularization parameters are reported in Tables~}\ref{tab:search-space-a}\hl{ and~}\ref{tab:search-space-b}\hl{ (Appendix).} \hl{The configurations these searches select are used by the feature-selection and feature-ranking analyses reported below, and were not re-selected for the reported campaign. To assess the effect of reusing them, we additionally re-tuned \textsc{catboost} and \textsc{xgboost}, the two models used by these analyses, on the reported campaign, with cross-validation folds grouped by simulation run, and compared the newly selected configurations with the inherited ones on the held-out test partition. The resulting differences do not exceed one accuracy point and are not consistent in direction across models or configurations. The classifier comparison does not use these configurations; its models are constructed with settings fixed in the detection pipeline, which prevents arm-specific hyperparameter tuning from contributing to differences between experimental arms.}

\hl{The linear \textsc{svm}'s regularization parameter was additionally swept over $C \in [10^{-3},\,10^{6}]$, with the iteration limit raised tenfold and the convergence tolerance held fixed across the grid. Selection was restricted to converged fits, giving $C=10^{-2}$ under static conditions and $C=1$ under mobility. Support vector machines do not produce probability estimates directly: the \textsc{rbf} variant obtains them through internal Platt scaling, whereas the linear variants use an isotonic calibration wrapper.}

\hl{The configurations selected by the unified search described above} are reported in Tables~\ref{tab:hyperparameters_static} and~\ref{tab:hyperparameters_mobile} (Appendix). \hl{Boosting models use a fixed iteration count without early stopping.}

\subsection{Threshold Selection and Calibration}
\label{subsec:threshold_calibration}

For each classifier, a decision threshold is selected \hl{on the threshold-selection half of the validation partition, from the resulting prediction scores,} by evaluating thresholds in $[0.1, 0.9]$ in steps of $0.01$ and choosing the value that maximizes \hl{accuracy on that half. Accuracy is used because the two classes are treated as equally important and the two types of misclassification are assigned equal cost, consistent with the reconnaissance setting studied here.}

\hl{For the individually calibrated classifiers, probability calibration is performed using isotonic regression fitted on the calibration half of the validation partition, with the training-fitted classifier held fixed}~\cite{zadrozny2002transforming,niculescu2005predicting}\hl{. The calibration and threshold-selection halves are disjoint and grouped by simulation run, so the pipeline proceeds from training to calibration, threshold selection, and finally the test partition without reusing observations between these stages. Calibration targets probability quality, whereas threshold selection defines the operating decision rule. The Stacking Ensemble is built from the uncalibrated models and has no calibrator in its prediction path; only its operating threshold is selected on the threshold-selection half.}

\hl{Two controls quantify distinct potential sources of bias. For the individually calibrated classifiers, a matched control isolating the effect of reusing calibration observations for threshold selection changes held-out accuracy by $-0.01$ percentage points on average under static conditions and $-0.06$ under mobility, with medians of $0.00$ and $-0.02$, respectively, providing no evidence of systematic inflation from such reuse. Separately, within the disjoint protocol used here, the residual optimism associated with selecting a threshold on a particular sample, estimated by cross-validation over the threshold-selection observations grouped by simulation run, is $+0.08$ percentage points for the Stacking Ensemble under static conditions and $+0.13$ under mobility.}

\hl{Threshold choice has only a minor effect under the equal-cost criterion. The accuracy-maximizing threshold computed post hoc on the test set itself, which uses test labels and is therefore reported only as an unattainable bound, exceeds accuracy at the selected operating point by $0.12$ percentage points under static conditions and $0.08$ under mobility. Every threshold in $[0.420, 0.650]$ under static conditions and $[0.380, 0.550]$ under mobility lies within $0.1$ percentage points of that bound. A cost-sensitive analysis under asymmetric misclassification costs is reported in Table~}\ref{tab:cost-sensitive}\hl{ (Appendix).}

Reported accuracy and $F_1$ are based on thresholded predictions, whereas \textsc{auc} is computed from \hl{continuous prediction scores} and is therefore independent of the decision threshold.

\subsection{Stacking Ensemble}
\label{subsec:stacking_ensemble}

A two-level stacking ensemble \cite{wolpert1992stacked} is constructed to combine the base classifiers. \hl{The base layer comprises ten of the sixteen configurations: the three tree-based ensembles, the six gradient boosting library configurations, and Bagging-RF. The remaining configurations, including the linear and kernel-based models, do not enter the meta-feature space.}

Meta-features are generated using out-of-fold predictions obtained via $5$-fold cross-validation on the training partition, \hl{preventing in-sample meta-features}. The meta-learner is a Logistic Regression model with L2 regularization ($C = 1.0$), trained solely on the base-model probabilities. Raw features are not passed to the meta-layer.

\hl{The ensemble is built from the uncalibrated base models, so no calibrator enters its prediction path. Its operating threshold is selected on the threshold-selection half by the same criterion as the base classifiers, giving $0.640$ under static conditions and $0.480$ under mobility. Relative to a fixed threshold of $0.5$, this changes test accuracy by $+0.01$ percentage points in both configurations.}

\subsection{Evaluation and Reproducibility}
\label{subsec:evaluation_reproducibility}

Performance is evaluated on the test partition, which is excluded from \hl{feature selection, model fitting, calibration, and threshold selection}. Because each simulation run contributes four correlated measurement windows, confidence intervals are estimated using cluster bootstrap resampling at the simulation-run level. Each bootstrap replicate samples runs with replacement and retains all associated windows, thereby preserving the within-run dependence structure. \hl{Each comparison names its own statistic and its own number of clusters.} The trained models, calibration procedure, preprocessing pipeline, and validation-selected decision thresholds remain fixed throughout. Reported confidence intervals are \hl{two-sided} $95\%$ percentile intervals based on $2{,}000$ bootstrap resamples.

\hl{Two campaign scales are used. The main campaign comprises $10{,}000$ accepted runs per configuration, so its test partition holds $2{,}000$ runs and $8{,}000$ measurement windows. The controlled experiments reported later use paired campaigns of $2{,}000$ runs, giving test partitions of $400$ runs and $1{,}600$ windows, and their confidence intervals are therefore expected to be approximately twice as wide. The \textsc{fpnt} evaluation follows the same procedure on its own campaign of $2{,}000$ runs per configuration: the partition, the calibration and threshold-selection halves, and the test partition are formed as above, and feature selection, model fitting, calibration, and threshold selection are carried out on those data rather than transferred from the \textsc{gcop}/\textsc{gcohp} campaign.}

Code and processed data used in this study are publicly available at
\url{https://github.com/nschwe/dcfm-defense-detection}.

All in-domain results correspond to a single deterministic pipeline execution. Cross-domain and feature stability analyses are evaluated over $20$ \hl{repetitions. Each repetition draws a separately randomized run-grouped $60/20/20$ partition of the same campaign, refits the classifier, re-selects the decision threshold on that repetition's threshold-selection data, and scores that repetition's test partition once. These are repeated random subsamplings of a fixed dataset rather than resampling with replacement, and quantities reported over them are given as means with the sample standard deviation across the twenty repetitions.}

\hl{Every inferential comparison in the paper is reported as a $95\%$ bootstrap confidence interval, and the resampling unit is the unit that generates the variability being quantified. Comparisons between arms are paired at the level of the simulation run, so the run is the resampling unit and the same resampled runs are applied to both arms, with the paired difference as the reported quantity. Quantities computed once per repetition vary from repetition to repetition, so there the repetition is the resampling unit. The unit is stated with each family of comparisons.}

\hl{Training and deployment cost were measured on the reported campaign. A complete execution of the reported training pipeline takes $8$~minutes $49$~seconds under static conditions and $9$~minutes $7$~seconds under mobility; this is an offline cost and is not incurred per window. Once trained, scoring a single measurement window with the stacking ensemble takes $174.4$~ms under static conditions and $175.9$~ms under mobility, against a $40$~s observation window. The serialised detector, comprising the ensemble, the scaler, the selected feature names and the decision threshold, occupies $712.9$~MB and $623.3$~MB on disk, and a process that loads it and nothing else holds $1{,}705.9$~MB and $1{,}527.5$~MB of resident memory while scoring windows. The detector scores one window per observation interval and takes no part in routing. A comparison over all seventeen classifiers, giving the accuracy, \textsc{auc}, serialised size and per-window latency of each, is released with the code and data as a deployment-cost frontier; the stacking ensemble remains the detector reported here.}

\section{Results}

\subsection{In-Domain Classification Performance}
\label{subsec:in-domain}

Table~\ref{tab:in_domain_performance} reports in-domain classification performance for \hl{eleven} representative models spanning standard model families: meta-ensemble, boosting, tree-based ensembles, linear, and kernel-based methods. The full evaluation covered \hl{the sixteen classifier configurations of Section~}\ref{subsec:classifiers-hyperparameters} \hl{and the Stacking Ensemble}; the selected models span the observed performance range. \hl{The two additional baselines of the same section are reported in the lower block of the table and are not among the seventeen.} Confidence intervals are $95\%$ cluster-bootstrap intervals computed at the simulation-run level, as described in Section~\ref{subsec:evaluation_reproducibility}.

\begin{table*}[t]
\centering
\scriptsize
\setlength{\tabcolsep}{4pt}

\caption{In-domain classification performance under static and mobile configurations. Accuracy values are reported with 95\% cluster-bootstrap confidence intervals (2{,}000 resamples at the simulation-run level; see Section~\ref{subsec:evaluation_reproducibility}). $n \approx 2{,}000$ test runs and $n \approx 8{,}000$ test windows per configuration. \hl{The lower block reports two additional baselines that are not among the seventeen; neither hinge-loss fit converged within the iteration limit.}}
\label{tab:in_domain_performance}

\begin{tabular}{@{}lcccc@{}}
\toprule
Model & Static Acc. & Mobile Acc. & Static AUC & Mobile AUC \\
\midrule
Stacking Ensemble                & 0.8998 [0.8918, 0.9078] & 0.9298 [0.9231, 0.9360] & 0.9403 & 0.9729 \\
\textsc{XGBoost} (conservative)  & 0.8989 [0.8906, 0.9070] & 0.9149 [0.9080, 0.9218] & 0.9414 & 0.9605 \\
\textsc{XGBoost} (balanced)      & 0.8985 [0.8900, 0.9068] & 0.9163 [0.9095, 0.9228] & 0.9370 & 0.9606 \\
\textsc{CatBoost} (balanced)     & 0.8971 [0.8890, 0.9054] & 0.9185 [0.9118, 0.9251] & 0.9406 & 0.9609 \\
\textsc{CatBoost} (deep)         & 0.8956 [0.8871, 0.9038] & 0.9189 [0.9120, 0.9254] & 0.9382 & 0.9645 \\
Random Forest                    & 0.8768 [0.8682, 0.8856] & 0.8986 [0.8914, 0.9059] & 0.9210 & 0.9454 \\
AdaBoost                         & 0.8955 [0.8874, 0.9035] & 0.9061 [0.8989, 0.9133] & 0.9382 & 0.9528 \\
Logistic Regression              & 0.8995 [0.8914, 0.9075] & 0.8805 [0.8725, 0.8883] & 0.9291 & 0.9176 \\
Ridge                            & 0.8751 [0.8666, 0.8835] & 0.8791 [0.8709, 0.8873] & 0.9188 & 0.9169 \\
\textsc{svm} (\textsc{rbf})      & 0.8843 [0.8760, 0.8926] & 0.9085 [0.9013, 0.9155] & 0.9219 & 0.9414 \\
\textsc{svm} (linear)            & 0.8935 [0.8853, 0.9014] & 0.8784 [0.8700, 0.8865] & 0.9243 & 0.9171 \\
\midrule
\textsc{lda}                     & 0.8746 [0.8661, 0.8830] & 0.8784 [0.8701, 0.8866] & 0.9188 & 0.9170 \\
Linear \textsc{svm} (hinge loss) & 0.8963 [0.8879, 0.9043] & 0.8870 [0.8791, 0.8946] & 0.9269 & 0.9214 \\
\bottomrule
\end{tabular}
\end{table*}

\begin{figure*}[t]
\centering
\includegraphics[width=\textwidth]{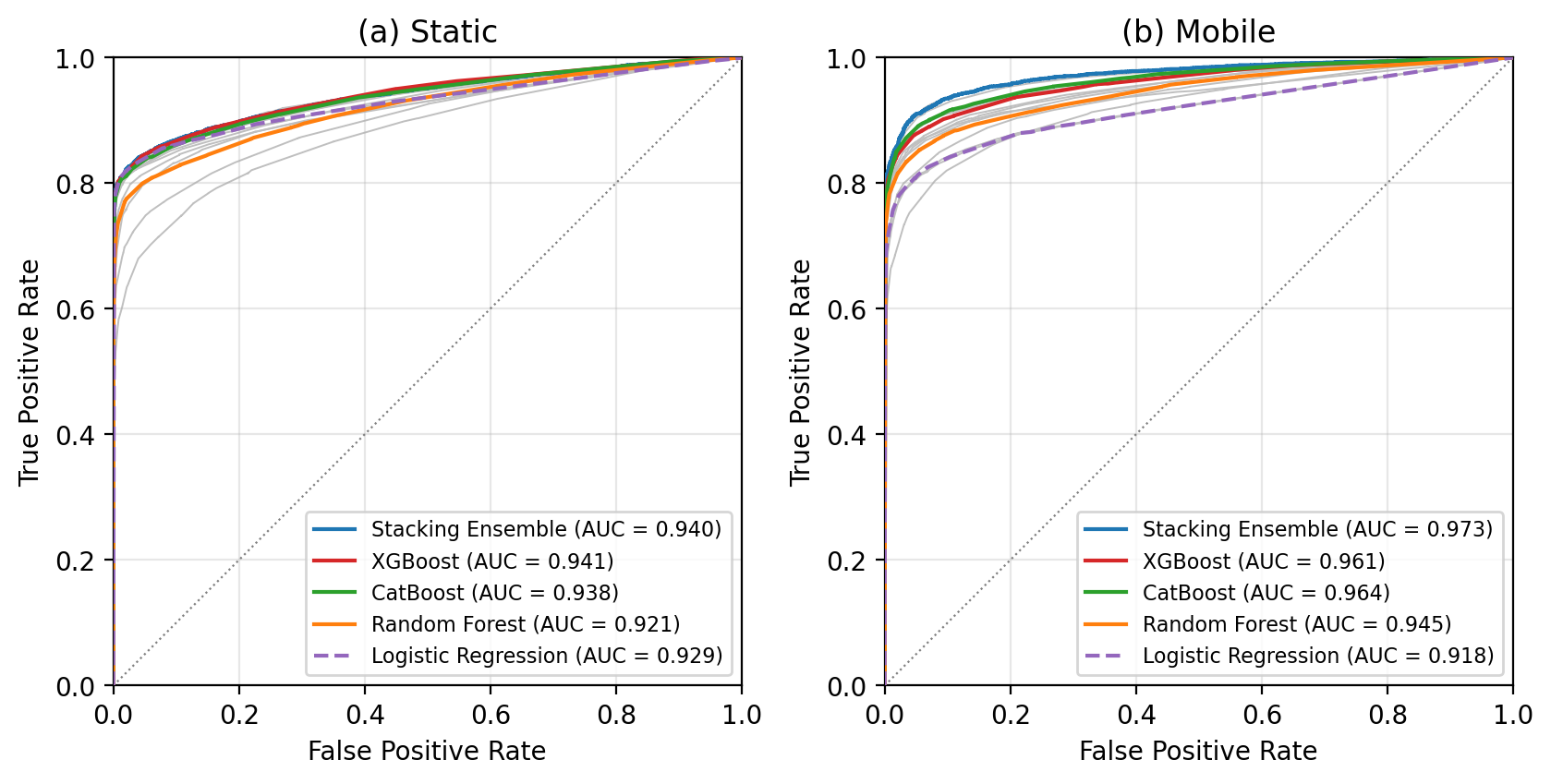}
\caption{\hl{\textsc{roc} curves for all seventeen models under static (a) and mobile (b) configurations. Five models are drawn in colour and named: the Stacking Ensemble, \textsc{xgboost}, \textsc{catboost}, Random Forest and Logistic Regression. The remaining twelve are drawn as light curves. \textsc{auc} ranges from \mbox{$0.8841$} (Bagging-ET) to \mbox{$0.9414$} under static conditions and from \mbox{$0.9169$} (Ridge) to \mbox{$0.9729$} under mobility.}}
\label{fig:roc}
\end{figure*}

\hl{Across the seventeen models, static accuracy ranges from \mbox{$0.8214$} to \mbox{$0.8998$} and \textsc{auc} from \mbox{$0.8841$} to \mbox{$0.9414$}; under
mobility accuracy ranges from \mbox{$0.8619$} to \mbox{$0.9298$} and \textsc{auc} from \mbox{$0.9169$} to \mbox{$0.9729$}. No evaluated model approaches chance
performance in either configuration.} The confidence intervals of the top models overlap.

Performance is \hl{higher under mobile conditions for every tree-based, boosting, and meta-ensemble model}. The Stacking Ensemble improves from \hl{\mbox{$0.8998$}} to \hl{\mbox{$0.9298$}} (\textsc{auc}: \hl{\mbox{$0.9403$}}$\to$\hl{\mbox{$0.9729$}}), \hl{and all thirteen evaluated models in these families gain between \mbox{$0.0074$} and \mbox{$0.0405$}
accuracy points.}

\hl{Among the tree-based, boosting and meta-ensemble models the} relative ranking is largely preserved across configurations. \hl{At the top of the \textsc{auc} ranking, \textsc{xgboost} (conservative) achieves the highest static value (\mbox{$0.9414$}), whereas the Stacking Ensemble attains the highest value under mobility (\mbox{$0.9729$}).}

The behavior of \hl{the linear and kernel-based} models under mobility is not uniform. \hl{Logistic Regression decreases from \mbox{$0.8995$} to \mbox{$0.8805$} and the linear \textsc{svm} from \mbox{$0.8935$} to \mbox{$0.8784$}, whereas Ridge is nearly unchanged (\mbox{$0.8751 \rightarrow 0.8791$}) and the kernel-based \textsc{svm} (\textsc{rbf}) improves from \mbox{$0.8843$} to \mbox{$0.9085$}. For the two additional baselines, Linear Discriminant Analysis attains \mbox{$0.8746$} and \mbox{$0.8784$}, while a linear \textsc{svm} with hinge loss attains \mbox{$0.8963$} and \mbox{$0.8870$}. Neither hinge-loss fit converged within the iteration limit, and raising the limit tenfold changed accuracy by less than \mbox{$0.001$}.}

\hl{The \textsc{roc} curves in Figure~}\ref{fig:roc} \hl{show the corresponding ranking performance across all seventeen models.}

\hl{The second evaluation applies the same single-vantage observation model and the same learning pipeline to \textsc{fpnt} under a black-hole attack, with feature selection, model fitting and threshold selection carried out independently on its own campaign. The stacking ensemble attains $0.9488$ accuracy (\textsc{auc} $0.9886$) under static conditions and $0.9394$ (\textsc{auc} $0.9855$) under mobility. On a size-matched \textsc{gcop}/\textsc{gcohp} campaign the corresponding figures are $0.8856$ (\textsc{auc} $0.9326$) and $0.9063$ (\textsc{auc} $0.9525$). Because the two defenses are evaluated against different attacks, these figures characterize each defense in its own setting and do not constitute a controlled comparison between them.}

\hl{Mean frame size is the most discriminative individual observable in the \textsc{fpnt} campaign, with a single-feature \textsc{auc} of $0.8759$ under static conditions and $0.8572$ under mobility, and three of the five strongest individual observables are byte-volume quantities. Removing all four byte-volume observables and repeating the evaluation on the same campaign leaves accuracy at $0.9369$ under static conditions and $0.9238$ under mobility, reductions of $1.19$ and $1.56$ percentage points. Mean frame size is not among the features selected for \textsc{gcop}/\textsc{gcohp}.}

\hl{The \textsc{gcop}/\textsc{gcohp} results above motivate the cross-domain analysis in the following sections.}

\subsection{Per-Feature Class Separability}
\label{subsec:per-feature-separability}

To quantify how individual features respond to the presence of the defense mechanism, we compute \hl{the magnitude of} Cohen's $d$ class separation for each
feature under both static and mobile configurations. Table~\ref{tab:per_feature_separability} reports \hl{the three features of the \textsc{Universal-3} subset identified in Section~}\ref{subsec:feature-consistency}. \hl{All three retain univariate separation in both configurations. \texttt{AverageAdvertisedLinksPerTCMessage} and \texttt{TcMessageRate} change only modestly between configurations, whereas \texttt{AdvertisedLinksPerKnownNode} decreases from \mbox{$1.1043$} to
\mbox{$0.5538$} under mobility.}

\begin{table}[t]
\centering
\begin{threeparttable}

\caption{\hl{Magnitude of Cohen's $d$ class separation for the \textsc{Universal-3} features under static and mobile configurations, $|d| = |\bar{x}_{1} - \bar{x}_{0}| / s_{\mathrm{pooled}}$. Class~1 comprises the scenarios in which the defense is deployed (defense-only and defense-vs-attack) and class~0 those in which it is not (baseline and attack-only). The quantity is unsigned: larger values indicate stronger univariate separation, and the table does not indicate which class has the larger feature value. $n = 40{,}000$ measurement windows per configuration, $20{,}000$ per class.}}
\label{tab:per_feature_separability}
\begin{tabular}{lcc}
\toprule
Feature & \hl{Static $|d|$} & \hl{Mobile $|d|$} \\
\midrule
AverageAdvertisedLinksPerTCMessage & 1.3192 & 1.3678 \\
AdvertisedLinksPerKnownNode        & 1.1043 & 0.5538 \\
TcMessageRate                      & 0.9075 & 0.8948 \\
\bottomrule
\end{tabular}
\end{threeparttable}
\end{table}

\subsection{\hl{Protocol-Level Verification of the Observable Footprint}}
\label{subsec:protocol-footprint}

\hl{To verify that the control-plane changes associated with the selected
observables are present independently of classifier behavior, we compare
the baseline and defense-only conditions within each accepted seed. The
paired runs use the same topology and contain no attack; differences are
formed within each seed and resampled at the seed level ($n=10{,}000$ per
configuration).}

\hl{All three measured quantities, computed before feature generation,
increase under the defense in both configurations. Advertised links per
received \textsc{tc} message rise by $0.88$ $[0.86, 0.89]$ under static
conditions and $0.94$ $[0.92, 0.95]$ under mobility, increases of $20.5$
and $22.9$ per cent; received \textsc{tc} messages increase by $245.1$
$[239.1, 251.2]$ and $403.4$ $[394.4, 412.6]$, or $39.7$ and $59.4$ per
cent; and received \textsc{olsr} control bytes increase by $12{,}885$
$[12{,}580, 13{,}197]$ and $19{,}641$ $[19{,}170, 20{,}123]$, or $29.9$
and $43.3$ per cent. None of the six intervals includes zero.}

\hl{The latter two quantities are raw observer-received counterparts of
retained features rather than the features themselves: received
\textsc{tc} messages replace the de-duplicated rate reported by
\mbox{\texttt{TcMessageRate}}, and control-byte volume replaces the
control-to-data ratio. \mbox{\texttt{AdvertisedLinksPerKnownNode}} is not
measured separately here.}

\subsection{Joint Feature Separability}
\label{subsec:joint_separability}

The in-domain performance gap between static and mobile configurations (Section~\ref{subsec:in-domain}) motivates examining whether the joint feature space is inherently more separable under mobility, or whether the observed performance gains are model-dependent. To address this, we evaluate two multivariate separability measures on the \hl{17} observable base features: the Mahalanobis distance between class centroids and the \textsc{lda} separation ratio.

All features are standardized to zero mean and unit variance using statistics computed on the full dataset of each configuration separately. The Mahalanobis distance is computed using a pooled covariance matrix estimated from both classes, with Tikhonov regularization ($\epsilon = 10^{-6}$) applied to ensure numerical stability in the presence of correlated features. The \textsc{lda} separation ratio is defined as the ratio of between-class to within-class variance along the Fisher discriminant direction. \hl{The difference between configurations is quantified by resampling whole simulation runs within each configuration, with all four measurement windows of a run kept together.}

\begin{table}[t]
\centering
\footnotesize
\caption{\hl{Multivariate separability measures for the $17$ observable base features under static and mobile configurations. Each configuration entry is the point estimate on its full dataset. The difference is accompanied by a $95\%$ cluster-bootstrap interval obtained by resampling whole simulation runs within each configuration ($B = 2{,}000$); the two configurations are separate campaigns and do not share runs, so the difference is between independent samples rather than paired. $n = 10{,}000$ simulation runs per configuration, contributing $40{,}000$ rows.}}
\label{tab:joint_separability}
\begin{tabular}{lccc}
\toprule
Measure & Static & Mobile & \hl{Difference (mobile $-$ static)} \\
\midrule
Mahalanobis distance          & 1.7472 & 1.6926 & $-0.0546$ $[-0.0995, -0.0077]$ \\
\textsc{lda} separation ratio & 0.7632 & 0.7162 & $-0.0470$ $[-0.0856, -0.0067]$ \\
\bottomrule
\end{tabular}
\end{table}

\hl{Both measures are lower under mobility} (Table~\ref{tab:joint_separability}). The Mahalanobis distance \hl{decreases} from \hl{\mbox{$1.7472$}} to \hl{\mbox{$1.6926$}}, and the \textsc{lda} separation ratio \hl{decreases} from \hl{\mbox{$0.7632$}} to \hl{\mbox{$0.7162$}}. \hl{The $95\%$ cluster-bootstrap intervals for both mobile-minus-static differences exclude zero.}

\subsection{Cross-Domain Generalization}
\label{subsec:cross-domain}

Cross-domain performance \hl{shows a limited} imbalance between transfer directions when using the full feature set (\hl{$K=17$})\hl{, where $K$ denotes the number of features in the subset used for classification}. Models trained on static data \hl{reach \mbox{$0.8753 \pm 0.0044$} when evaluated on mobile data}
(Table~\ref{tab:optimal-k}), whereas models trained on mobile data \hl{reach \mbox{$0.8522 \pm 0.0049$} on static data, a difference between transfer directions of \mbox{$0.0231$}. Relative to the corresponding in-domain results of the same sweep, \mbox{$0.9003$} and \mbox{$0.9211$}, transferring reduces accuracy by \mbox{$0.0251$} points in the static-to-mobile direction and by \mbox{$0.0689$} points in the reverse direction.}

These results \hl{motivate further analysis of how cross-domain performance depends on the number of features used}. In particular, the following sections examine whether a compact feature subset can \hl{preserve this transfer behavior} while maintaining competitive classification performance.

\subsection{Feature Consistency Across Configurations}
\label{subsec:feature-consistency}

To identify features that remain informative across both static and mobile configurations, we evaluated feature importance under six complementary criteria: native importance from Random Forest, XGBoost, and CatBoost, and permutation importance computed on each of these classifiers. All feature-importance analyses were conducted on the shared observable metric space used for cross-domain transfer evaluation. For each criterion and each configuration, we performed \hl{repeated run-grouped resampling ($20$ repetitions, $80\%$ stability threshold)} and recorded the resulting feature ranking. \hl{$K_{\text{rank}}$ denotes the depth of this per-configuration feature ranking.} A feature was considered stable under a given importance criterion only if it appeared in the corresponding top-$K_{\text{rank}}$ intersection in at least $80\%$ of \hl{repetitions} ($\geq 16$ out of $20$). This procedure mitigates the known instability of tree-based importance scores, particularly in the presence of correlated features.

\hl{For each importance criterion, static and mobile data produce separate feature rankings. At depth $K_{\text{rank}}$, we take the top $K_{\text{rank}}$ features from each ranking, in each repetition, and retain their intersection. Thus, $K_{\text{rank}}$ specifies how far down the two rankings are examined, not how many features are ultimately supplied to the classifier. The latter is denoted by $K$ and is selected separately in Section~}\ref{subsec:feature-size}\hl{.}

\hl{For each criterion, we swept $K_{\text{rank}}=1,\ldots,17$ and selected the smallest depth whose intersection reached at least $99\%$ of that criterion's highest mean cross-domain accuracy across both transfer directions. All six criteria select $K_{\text{rank}}=4$, and each yields a four-feature intersection at that depth. Three features are common to all six intersections:}

\begin{itemize}
  \item \texttt{AverageAdvertisedLinksPerTCMessage} -- advertised neighbor-link
        structure in \textsc{tc} messages,
  \item \hl{\mbox{\texttt{AdvertisedLinksPerKnownNode}} -- the sum of each
        originator's latest advertised-link count, normalized by the number of
        distinct node addresses observed,}
  \item \hl{\mbox{\texttt{TcMessageRate}} -- the rate of distinct
        topology-control messages received by the observer.}
\end{itemize}

\hl{We refer to this common set as \textsc{Universal-3}. The fourth feature is \mbox{\texttt{AverageHopCount}} for four criteria (\mbox{\texttt{catboost\_pvc}} and the three permutation criteria) and \mbox{\texttt{MeanSniffedFrameSize}} for the remaining two (native Random Forest and XGBoost importance). \mbox{\texttt{MeanSniffedFrameSize}} is the mean size of a frame received at the observer and is defined, together with every other observable and derived column, in Table~}\ref{tab:feature-list}\hl{. Table~}\ref{tab:importance_recommended_k}\hl{ reports these results and the corresponding cross-domain accuracies.}

\hl{This agreement emerges specifically at depth $4$. Only one feature is common to all six criteria at $K_{\text{rank}}=1$, $2$, and $3$; the number rises to three at $K_{\text{rank}}=4$, remains three at $K_{\text{rank}}=5$, and then grows gradually to the trivial full-set intersection at $K_{\text{rank}}=17$. Under native Random Forest importance, for example, static and mobile data contain the same four features within their top four but order them differently: each places a different one of \mbox{\texttt{AdvertisedLinksPerKnownNode}} and \mbox{\texttt{TcMessageRate}} outside its top three. A top-$3$ intersection therefore excludes both, whereas a top-$4$ intersection retains both.}

\hl{Section~}\ref{subsec:feature-size}\hl{ separately selects the number of features supplied to the classifier, under a different rule: a one-standard-deviation tolerance around the best mean cross-domain accuracy. It selects $K=3$, corresponding exactly to \textsc{Universal-3}; within the four-feature \mbox{\texttt{catboost\_pvc}} intersection used to anchor that analysis, \mbox{\texttt{AverageHopCount}} ranks fourth jointly across the two configurations.}

\hl{Because \textsc{Universal-3} is derived from agreement between the two configurations, we also repeated the selection using only the source configuration, with feature ranking computed on the source training partition alone. This control selects $K=3$ and recovers \textsc{Universal-3} in both transfer directions. The resulting zero-shot transfer accuracy is $0.8736 \pm 0.0063$ with \textsc{auc} $0.9125$ for \textsc{static}~$\rightarrow$~\textsc{mobile}, and $0.8563 \pm 0.0090$ with \textsc{auc} $0.9101$ for \textsc{mobile}~$\rightarrow$~\textsc{static}. This is a control on the selection procedure and not a substitute for it: the reported subset remains the one derived from both configurations. As a separate control, selection based on in-domain classification accuracy alone, independently within each configuration, also recovered \textsc{Universal-3}.}

\hl{Two limitations of this evidence should be stated. The six criteria are three tree-based learners under two importance definitions rather than six independent learning algorithms, so their agreement bounds the influence of the importance definition more tightly than that of the model family. And the source-only control removes target information from the ranking, but both configurations are drawn from the same simulator and traffic model, so it does not establish that the subset transfers to a different network generator.}

\begin{table*}[!ht]
\centering
\footnotesize
\caption{\hl{Recommended ranking depth $K_{\text{rank}}$ per importance criterion, with the feature that distinguishes each criterion's set and the cross-domain accuracy at that depth in each transfer direction. Every criterion returns \textsc{Universal-3} together with one additional feature; only that fourth feature is shown. Feature rankings are recomputed in each of $20$ repetitions and a feature is retained only if it appears in the static/mobile intersection in at least $16$ of them. Accuracies are means over the $20$ repetitions and over the three tree-based classifiers, evaluated on the full target configuration ($n = 40{,}000$ windows per direction).}}
\label{tab:importance_recommended_k}
\begin{tabular}{lcccc}
\toprule
Importance criterion & $K_{\text{rank}}$ & Fourth feature & S$\rightarrow$M & M$\rightarrow$S \\
\midrule
\texttt{catboost\_pvc}  & 4 & \texttt{AverageHopCount} & 0.8786 & 0.8590 \\
\texttt{perm\_catboost} & 4 & \texttt{AverageHopCount} & 0.8786 & 0.8590 \\
\texttt{perm\_rf}       & 4 & \texttt{AverageHopCount} & 0.8786 & 0.8590 \\
\texttt{perm\_xgb}      & 4 & \texttt{AverageHopCount} & 0.8786 & 0.8590 \\
\texttt{rf\_native}     & 4 & \texttt{MeanSniffedFrameSize} & 0.8790 & 0.8561 \\
\texttt{xgb\_gain}      & 4 & \texttt{MeanSniffedFrameSize} & 0.8790 & 0.8561 \\
\bottomrule
\end{tabular}
\end{table*}

\subsection{Effect of Feature Set Size}
\label{subsec:feature-size}

To characterize how cross-domain transfer depends on the number of features used, we evaluated CatBoost classifiers trained on feature subsets of varying size $K$, \hl{for \mbox{$K = 1, \ldots, 17$}. The sweep is anchored on the four-feature intersection obtained under \mbox{\texttt{catboost\_pvc}} at \mbox{$K_{\text{rank}} = 4$} in Section~}\ref{subsec:feature-consistency}\hl{. For \mbox{$K < 4$}, only the $K$ anchor features ranked best on average across the two configurations are retained; for \mbox{$K > 4$}, the next \mbox{$(K-4)$} features are appended in order of average \mbox{\texttt{catboost\_pvc}} importance across the two configurations.} For each $K$, we trained the classifier on the source configuration (\textsc{static} or \textsc{mobile}) and evaluated it on both the source and the target configurations. The procedure follows the train/validation/test split of Section~\ref{subsec:data_partitioning}, with the decision threshold tuned on the source validation partition. \hl{In-domain scores use the held-out source test partition (\mbox{$n = 8{,}000$} windows), because the model was fitted on that configuration; cross-domain scores use the full target configuration (\mbox{$n = 40{,}000$} windows), none of which was used for model fitting or selection. Results are averaged over $20$ repetitions.}

\begin{table*}[!ht]
\centering
\footnotesize
\caption{\hl{Cross-domain performance as a function of feature set size $K$, with CatBoost (tuned hyperparameters) trained on the source configuration and evaluated on the target. The sweep is anchored on the four-feature intersection of Section~}\ref{subsec:feature-consistency}\hl{; for \mbox{$K < 4$} features are dropped from that anchor and for \mbox{$K > 4$} features are appended, both by \mbox{\texttt{catboost\_pvc}} importance. Accuracy values are mean \mbox{$\pm$} sample standard deviation over $20$ repetitions, and asymmetry is the difference between the two accuracy columns. Every entry is evaluated on the full target configuration (\mbox{$n = 40{,}000$} windows per transfer direction). The sweep covered \mbox{$K = 1, \ldots, 17$}; the table reports a representative subset of those sizes.}}
\label{tab:optimal-k}
\begin{tabular}{cccccc}
\toprule
$K$ & S$\rightarrow$M & M$\rightarrow$S & AUC (S$\rightarrow$M) & AUC (M$\rightarrow$S) & Asymmetry \\
\midrule
1  & $0.7889 \pm 0.0088$ & $0.7502 \pm 0.0027$ & 0.8518 & 0.8233 & 0.0387 \\
2  & $0.7580 \pm 0.0176$ & $0.8045 \pm 0.0016$ & 0.8712 & 0.8547 & 0.0465 \\
$\mathbf{3}$ & $\mathbf{0.8730 \pm 0.0053}$ & $\mathbf{0.8632 \pm 0.0082}$ & $\mathbf{0.9130}$ & $\mathbf{0.9108}$ & $\mathbf{0.0098}$ \\
5  & $0.8784 \pm 0.0048$ & $0.8503 \pm 0.0081$ & 0.9160 & 0.9146 & 0.0281 \\
9  & $0.8716 \pm 0.0051$ & $0.8557 \pm 0.0053$ & 0.9132 & 0.9108 & 0.0159 \\
13 & $0.8748 \pm 0.0048$ & $0.8526 \pm 0.0034$ & 0.9134 & 0.9092 & 0.0222 \\
17 & $0.8753 \pm 0.0044$ & $0.8522 \pm 0.0049$ & 0.9129 & 0.9085 & 0.0231 \\
\bottomrule
\end{tabular}
\end{table*}

\begin{figure}[t]
\centering
\includegraphics[width=\linewidth]{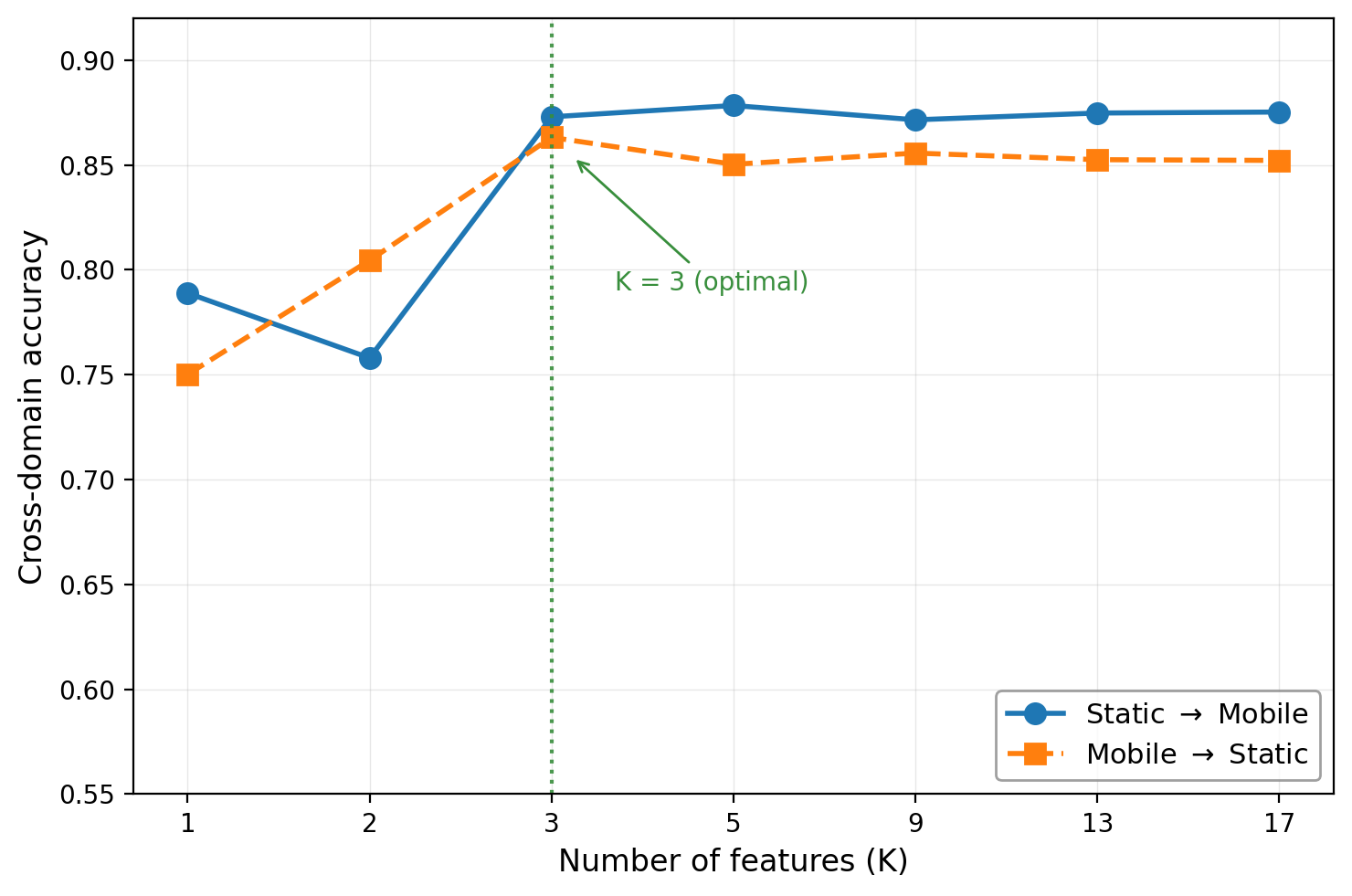}
\caption{\hl{Cross-domain accuracy as a function of feature set size ($K$). The two transfer directions come closest to agreement at \mbox{$K=3$} (\textsc{Universal-3}), where the gap between them is $0.0098$; below \mbox{$K=3$} both directions are markedly weaker, and above it neither direction improves materially. Standard deviations over $20$ repetitions are reported in Table~}\ref{tab:optimal-k}\hl{. The sweep covered \mbox{$K = 1, \ldots, 17$}; the figure and Table~}\ref{tab:optimal-k}\hl{ report a representative subset, and the \mbox{$x$}-axis is accordingly non-uniform.}}
\label{fig:optimal-k}
\end{figure}

Table~\ref{tab:optimal-k} and Figure~\ref{fig:optimal-k} report cross-domain accuracy and \textsc{auc} as a function of $K$. \hl{Both transfer directions are substantially weaker below \mbox{$K=3$}: \textsc{static}~$\rightarrow$~\textsc{mobile} accuracy is $0.7889$ at \mbox{$K=1$} and $0.7580$ at \mbox{$K=2$}, while \textsc{mobile}~$\rightarrow$~\textsc{static} moves in the opposite direction over the same step, from $0.7502$ to $0.8045$. At \mbox{$K=3$} both directions rise together, to $0.8730$ and $0.8632$. Beyond \mbox{$K=3$} both directions stay within a narrow band: \textsc{static}~$\rightarrow$~\textsc{mobile} varies between $0.8716$ and $0.8807$ and \textsc{mobile}~$\rightarrow$~\textsc{static} between $0.8503$ and $0.8563$ over the remaining fourteen sizes, and at the full observable set (\mbox{$K=17$}) they stand at $0.8753$ and $0.8522$.}

\hl{At \mbox{$K=3$} the two transfer directions come closest to agreement (\textsc{static}~$\rightarrow$~\textsc{mobile}: $0.8730$, \textsc{mobile}~$\rightarrow$~\textsc{static}: $0.8632$, asymmetry: $0.0098$). No other size in the sweep is within $0.0060$ of that figure: the next smallest asymmetry is $0.0159$ at \mbox{$K=9$}, and the remaining sizes lie between $0.0199$ and $0.0465$. The additional features introduced beyond the third therefore provide no material overall improvement in cross-domain transfer. They produce only modest changes in either direction, and they do not help in-domain either: source-configuration accuracy falls slightly from $0.9032$ to $0.9003$ under \textsc{static} and from $0.9291$ to $0.9211$ under \textsc{mobile} as $K$ grows from $3$ to $17$.}

\hl{\textsc{auc} follows the same pattern and is likewise flat above \mbox{$K=3$}: $0.9130$ and $0.9108$ at \mbox{$K=3$} against $0.9129$ and $0.9085$ at \mbox{$K=17$}. Ranking quality therefore shows no material improvement beyond three features. Among the sizes whose mean cross-domain accuracy lies within a one-SD tolerance around the best, where the tolerance is the mean of the two directional standard deviations at the best-performing size, \mbox{$K = 3$} is the smallest. It is also the size at which the two transfer directions agree most closely, with an asymmetry of $0.0098$ against $0.0199$ for the next best in that set. Accordingly, mean cross-domain accuracy is used to select $K$, while asymmetry characterizes the balance between the two transfer directions.}

\begin{figure}[t!]
\centering
\includegraphics[
width=\linewidth,
trim={0 8mm 0 0},
clip
]{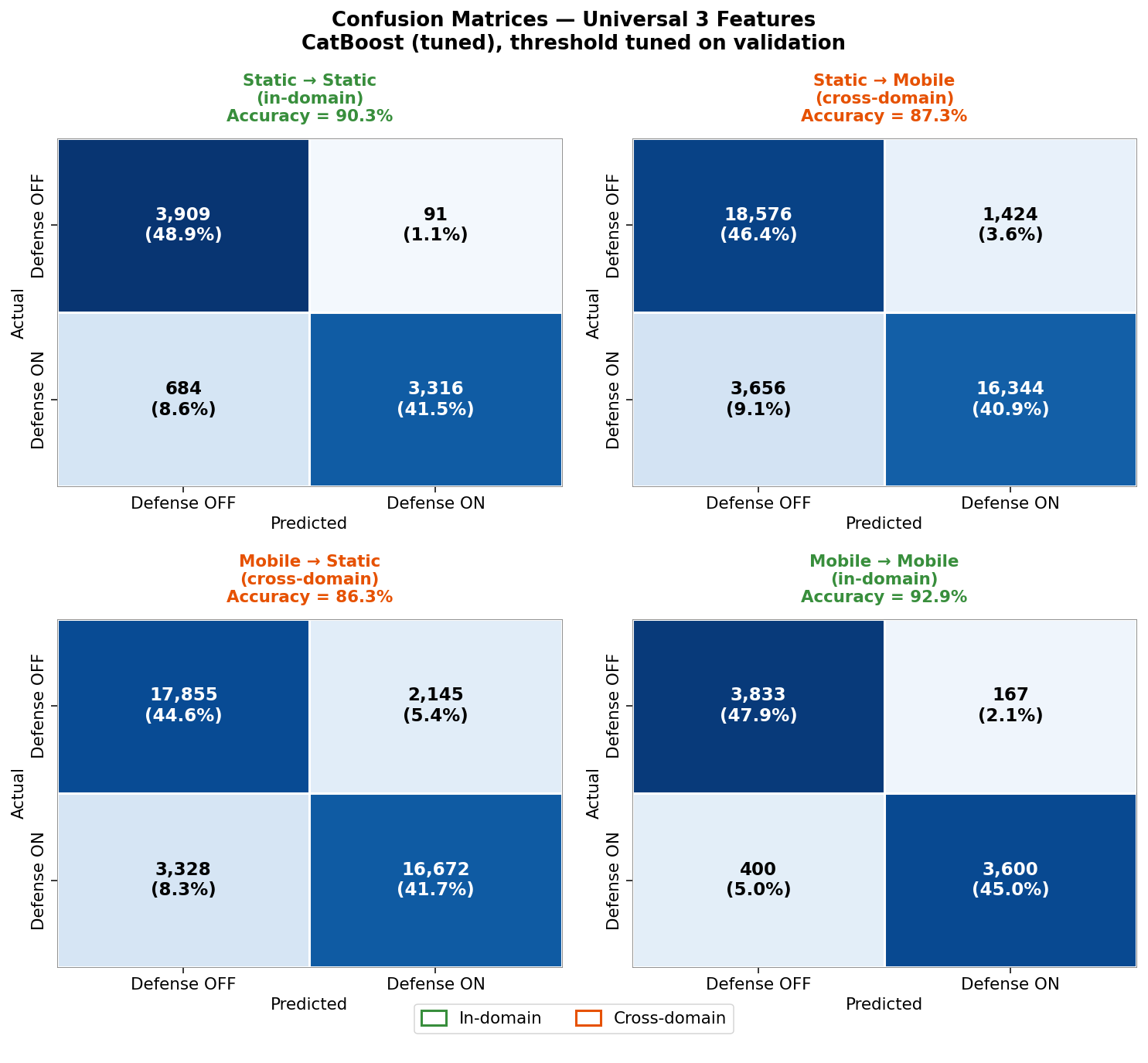}

\caption{\hl{Confusion matrices for \textsc{Universal-3} (\mbox{$K=3$}) under all four configuration pairs: in-domain (\textsc{static}~$\rightarrow$~\textsc{static}, \textsc{mobile}~$\rightarrow$~\textsc{mobile}) and cross-domain (\textsc{static}~$\rightarrow$~\textsc{mobile}, \textsc{mobile}~$\rightarrow$~\textsc{static}). In-domain panels are evaluated on the held-out test partition (\mbox{$8{,}000$} samples); cross-domain panels on the full target configuration (\mbox{$40{,}000$} samples), which the model never observes during training. Cells display sample counts and percentages of the panel total, averaged over $20$ repetitions.}}
\label{fig:confusion-universal}
\end{figure}

Figure~\ref{fig:confusion-universal} presents the in-domain and cross-domain confusion matrices for \hl{\textsc{Universal-3}. In-domain performance is preserved despite the reduction from $17$ observables to $3$ features ($0.9032$ for \textsc{static}, $0.9291$ for \textsc{mobile}), and matches the full pipeline: three features with CatBoost against seventeen observables with the stacking ensemble give $0.9032$ against $0.8998$ under \textsc{static} and $0.9291$ against $0.9298$ under \textsc{mobile}. Cross-domain transfer remains nearly symmetric in both directions ($0.8730$ for \textsc{static}~$\rightarrow$~\textsc{mobile}, $0.8632$ for \textsc{mobile}~$\rightarrow$~\textsc{static}), matching the \mbox{$K=3$} row of Table~}\ref{tab:optimal-k}\hl{ exactly.}

\hl{The same analysis was repeated with the Gauss-Markov campaign of Section~}\ref{sec:simulation-environment}\hl{ in place of the mobile configuration. The selected observables largely hold: at \mbox{$K = 3$}, eight of the twelve combinations of importance criterion and set-construction rule return exactly \textsc{Universal-3}. The twelve combinations comprise the six importance criteria under two rules: one retains only observables passing the stability threshold, whereas the other fills the set to \mbox{$K$} by mean importance. \texttt{AverageAdvertisedLinksPerTCMessage} is returned by all twelve. Detection within the Gauss-Markov configuration is also strong, $0.9339$. Transfer, however, is strongly asymmetric: training on the static configuration gives $0.8330$ under Gauss-Markov, whereas training under Gauss-Markov gives $0.5610$ on static data, an asymmetry of $0.2720$ against the $0.0098$ of Table~}\ref{tab:optimal-k}\hl{. Applied to this pair, the selection rule returns \mbox{$K = 1$}; the figures above are reported at \mbox{$K = 3$} to permit a like-for-like comparison with the main static and random-walk analysis. The near-symmetric transfer reported above therefore does not carry over to a different mobility process, and the cross-domain result of this work is stated for the static and random-walk configurations it evaluates.}

\subsection{\hl{Feature Instability and Cross-Domain Transfer}}
\label{subsec:feature-ablation-and-augmentation}

Sections~\ref{subsec:feature-consistency} and~\ref{subsec:feature-size} \hl{showed that increasing the feature set beyond the three \textsc{Universal-3} observables provides no material overall improvement in cross-domain transfer. We therefore ask whether observables that do not help can be identified in advance from cross-configuration instability. If instability predicted transfer harm, removing the most unstable observables would improve transfer, providing a simple feature-pruning criterion.}

\hl{We grouped the seventeen observables into five measurement-based families that partition them and removed each family in turn, retraining and re-evaluating from scratch. Two nested subsets of the control-plane family were included as controls. We evaluated CatBoost and Logistic Regression in both transfer directions. Instability is the mean absolute change in the \emph{magnitude} of per-feature Cohen's $d$ between configurations over the removed observables, denoted $\Delta|d|$. Results are based on $20$ repetitions and cross-domain accuracy is evaluated on the full target configuration ($n = 40{,}000$ windows). Additional controls remove individual observables or two randomly chosen
observables.}

\begin{table}[H]
\centering
\footnotesize
\caption{\hl{Change in cross-domain accuracy when a group of observables is removed, against the full seventeen-observable baseline. $\Delta|d|$ is the mean absolute change in the magnitude of per-feature Cohen's $d$ between configurations over the removed observables; it is the instability measure being tested and is never used to choose what to remove. Each accuracy $\Delta$ is the mean of $20$ paired repetition-level differences against the baseline, with a percentile bootstrap interval over those differences ($B = 10{,}000$), evaluated on the full target configuration ($n = 40{,}000$ windows per direction). The first five groups partition the seventeen observables; the last two are nested subsets of the control-plane group, reported as controls.}}
\label{tab:instability-controls}
\sisetup{
  table-number-alignment = center,
  table-format = +1.4,
  retain-explicit-plus = true
}
\begin{tabular}{lcllS[table-format=+1.4]@{\hspace{5pt}}l}
\toprule
Removed group & $\Delta|d|$ & Classifier & Dir. & {$\Delta$} & 95\% CI \\
\midrule
Control plane (6) & 0.1035 & CatBoost & S$\rightarrow$M & -0.1643 & [$-0.1704$, $-0.1586$] \\
                  &        &          & M$\rightarrow$S & -0.1016 & [$-0.1048$, $-0.0983$] \\
                  &        & LogReg   & S$\rightarrow$M & -0.2045 & [$-0.2099$, $-0.1987$] \\
                  &        &          & M$\rightarrow$S & -0.0032 & [$-0.0069$, $+0.0010$] \\
\addlinespace
Control overhead (2) & 0.1906 & CatBoost & S$\rightarrow$M & +0.0004 & [$-0.0011$, $+0.0019$] \\
                  &        &          & M$\rightarrow$S & +0.0007 & [$-0.0026$, $+0.0040$] \\
                  &        & LogReg   & S$\rightarrow$M & -0.0282 & [$-0.0321$, $-0.0245$] \\
                  &        &          & M$\rightarrow$S & +0.0038 & [$+0.0002$, $+0.0083$] \\
\addlinespace
Source visibility (2) & 0.2005 & CatBoost & S$\rightarrow$M & +0.0050 & [$+0.0035$, $+0.0066$] \\
                  &        &          & M$\rightarrow$S & -0.0030 & [$-0.0059$, $-0.0005$] \\
                  &        & LogReg   & S$\rightarrow$M & +0.0776 & [$+0.0719$, $+0.0835$] \\
                  &        &          & M$\rightarrow$S & +0.0055 & [$+0.0019$, $+0.0095$] \\
\addlinespace
Frame size (1) & 0.1614 & CatBoost & S$\rightarrow$M & -0.0016 & [$-0.0038$, $+0.0005$] \\
                  &        &          & M$\rightarrow$S & +0.0013 & [$-0.0015$, $+0.0042$] \\
                  &        & LogReg   & S$\rightarrow$M & -0.0211 & [$-0.0234$, $-0.0188$] \\
                  &        &          & M$\rightarrow$S & -0.0171 & [$-0.0226$, $-0.0110$] \\
\addlinespace
Traffic volume (6) & 0.0947 & CatBoost & S$\rightarrow$M & -0.0064 & [$-0.0091$, $-0.0037$] \\
                  &        &          & M$\rightarrow$S & -0.0043 & [$-0.0085$, $-0.0006$] \\
                  &        & LogReg   & S$\rightarrow$M & +0.0381 & [$+0.0350$, $+0.0411$] \\
                  &        &          & M$\rightarrow$S & -0.0431 & [$-0.0508$, $-0.0347$] \\
\midrule
\multicolumn{6}{l}{\emph{Nested controls within the control-plane group}} \\
\addlinespace
Control plane, protected (3) & 0.0030 & CatBoost & S$\rightarrow$M & -0.0035 & [$-0.0058$, $-0.0013$] \\
                  &        &          & M$\rightarrow$S & -0.0125 & [$-0.0159$, $-0.0090$] \\
                  &        & LogReg   & S$\rightarrow$M & -0.0008 & [$-0.0025$, $+0.0009$] \\
                  &        &          & M$\rightarrow$S & -0.0172 & [$-0.0227$, $-0.0117$] \\
\addlinespace
Constant columns (2) & 0.0000 & CatBoost & S$\rightarrow$M & 0.0000 & [\ \ \ $0.0000$, \ \ $0.0000$] \\
                  &        &          & M$\rightarrow$S & 0.0000 & [\ \ \ $0.0000$, \ \ $0.0000$] \\
                  &        & LogReg   & S$\rightarrow$M & {$-0.0000$} & [$-0.0007$, $+0.0009$] \\
                  &        &          & M$\rightarrow$S & -0.0009 & [$-0.0024$, $+0.0001$] \\
\bottomrule
\end{tabular}
\end{table}

\hl{The two groups with the largest $\Delta|d|$, source visibility ($0.2005$) and control overhead ($0.1906$), produce different removal effects. For \textsc{static}~$\rightarrow$~\textsc{mobile} Logistic Regression, removing source visibility changes accuracy by $+0.0776$, whereas removing control overhead changes it by $-0.0282$. The protected control-plane subset has $\Delta|d|=0.0030$; its removal changes CatBoost accuracy by $-0.0125$ in the \textsc{mobile}~$\rightarrow$~\textsc{static} direction. Removing the two constant columns produces no CatBoost change.}

\hl{At the individual-observable level, \texttt{AdvertisedLinksPerKnownNode} has the largest instability ($\Delta|d|=0.5505$) and belongs to \textsc{Universal-3}. Removing it produces losses of $-0.1526$ [$-0.1584$, $-0.1470$] for Logistic Regression and $-0.0727$ [$-0.0785$, $-0.0669$] for CatBoost in the \textsc{static}~$\rightarrow$~\textsc{mobile} direction; the intervals exclude zero in all four classifier--direction combinations.}

\hl{Removal effects also vary by classifier and transfer direction. Removing traffic-volume observables changes \textsc{static}~$\rightarrow$~\textsc{mobile} Logistic Regression by $+0.0381$ and \textsc{mobile}~$\rightarrow$~\textsc{static} by $-0.0431$, while the CatBoost changes remain below $0.0065$ in magnitude in both directions. Among the single-observable controls, removing any non-selected observable changes accuracy by at most $0.0276$, and by less than $0.0102$ except for \texttt{SourceVisibilityWindowStd}; for each of the two distinct randomly drawn pairs, the CatBoost intervals contain zero in both directions.}

\subsection{\hl{Sensitivity to Propagation and Medium-Access Assumptions}}
\label{subsec:robustness}

\hl{The evaluation above uses the range-based propagation model of Table~}\ref{tab:simulation-parameters}\hl{, in which every node inside the transmission range receives at full power and no node outside it receives anything. That idealization bears on this analysis more directly than it would on a performance study. Every observable the detector uses is a count or a rate formed from frames that actually reached the listener, so a model that settles reception by distance alone also settles, by construction, how much of the control plane a single vantage point observes. Two paired experiments therefore test how far the reported result depends on that assumption and on the medium-access conditions that accompany it. Both reuse the observation model and the pipeline of Sections~}\ref{subsec:data_partitioning}\hl{ to~}\ref{subsec:evaluation_reproducibility}\hl{ without modification, and in both, feature selection, model fitting and threshold selection are carried out independently within each arm, so each comparison is between two separately trained detectors rather than one detector transferred across settings.}

\hl{\textbf{Propagation.} We repeated the campaign with log-distance path loss at exponent $3.0$ and Nakagami fading, on paired seeds, leaving the traffic profile, node count, deployment area and physical-layer rate unchanged. Fading alters the connectivity graph, and no single calibration reproduces both properties of the range-based topology that matter here, so two operating points are reported. Matching the mean neighbor degree preserves the topology statistic but destroys route viability: only $18\%$ of static and $21\%$ of mobile deployments retain a viable source--victim route, against $61\%$ in both configurations at the second operating point. Calibrating instead to preserve route viability retains a comparable experimental population, at the cost of a denser effective graph. Both operating points are reported because each preserves one property of the range-based topology and gives up the other, and presenting only the one at which the network survives would amount to choosing the operating point by its outcome.}

\hl{\textbf{Medium access.} The reported campaign leaves the RTS/CTS handshake disabled, so hidden-terminal collisions are present in it. We therefore compared it against a paired configuration in which RTS/CTS is enabled to mitigate those collisions. RTS/CTS is the intervention here, not the configuration the rest of the paper uses.}

\begin{table}[!ht]
\centering
\footnotesize
\caption{\hl{Paired accuracy differences for the two sensitivity experiments, each against its own baseline. Each $\Delta$ is the difference between two separately trained detectors evaluated on the same simulation runs, with a $95\%$ percentile interval from $2{,}000$ cluster bootstrap resamples at the simulation-run level (Section~}\ref{subsec:evaluation_reproducibility}\hl{). Both arms of every comparison are scored with the stacking ensemble. $n$ is the test partition of a paired $2{,}000$-run campaign and is given per row, because the admission criterion removes different numbers of runs at the two propagation calibrations.}}
\label{tab:sensitivity}
\sisetup{
  table-number-alignment = center,
  table-format = +1.4,
  retain-explicit-plus = true
}
\begin{tabular}{llS[table-format=+1.4]@{\hspace{5pt}}lr}
\toprule
Experiment & Config. & {$\Delta$} & 95\% CI & $n$ (runs / windows) \\
\midrule
Faded, route viability & \textsc{static} & +0.0244 & [$-0.0006$, $+0.0487$] & 400 / 1{,}600 \\
                       & \textsc{mobile} & +0.0075 & [$-0.0169$, $+0.0313$] & 400 / 1{,}600 \\
\addlinespace
Faded, degree matched  & \textsc{static} & -0.3396 & [$-0.3697$, $-0.3068$] & 374 / 1{,}496 \\
                       & \textsc{mobile} & -0.3381 & [$-0.3663$, $-0.3081$] & 400 / 1{,}600 \\
\addlinespace
RTS/CTS enabled        & \textsc{static} & -0.0100 & [$-0.0206$, $0.0000$]  & 400 / 1{,}600 \\
                       & \textsc{mobile} & -0.0044 & [$-0.0137$, $+0.0050$] & 400 / 1{,}600 \\
\bottomrule
\end{tabular}
\end{table}

\hl{At the route-viability calibration, detection accuracy is $0.9094$ under static conditions and $0.9069$ under mobility, against $0.8850$ and $0.8994$ for the paired range-based baseline. Both intervals contain zero: we find no evidence that the less idealized propagation model degrades detection, and we do not claim that it improves it. At the degree-matched calibration accuracy falls to $0.5461$ and $0.5588$, close to chance. That endpoint bounds the experiment rather than measuring the detector: with four fifths of the deployments no longer carrying a viable source--victim route, there is little left to detect.}

\hl{Enabling RTS/CTS moves accuracy from $0.8944$ to $0.8844$ under static conditions and from $0.8956$ to $0.8913$ under mobility. The static interval runs from $-0.0206$ to zero and the mobile interval spans zero, so neither excludes it, and both point estimates are negative.}

\hl{One property of the route-viability calibration bounds what the propagation experiment shows. Preserving route viability under fading requires a more permissive loss threshold, and the resulting graph carries a mean neighbor degree $40$ to $52\%$ above the range-based model, the larger figure under static conditions. Detection at that operating point is therefore measured on a network richer in observable control traffic than the one the main campaign uses.}

\hl{Neither sensitivity experiment varies deployment density as an independently controlled factor. In particular, the route-viability calibration is not a density-robustness experiment because its effective graph is denser than that of the range-based baseline.}

\section{Discussion}
\label{sec:discussion}

\subsection{\hl{Why the Defense Is Observable}}
\label{subsec:mechanism}

\hl{The three features of \textsc{Universal-3} have a common protocol-level interpretation. All are computed from the Topology Control traffic the listener overhears, through which the defense operates: a defending node advertises a fictitious symmetric neighbor and carries that neighbor in the link set of the \textsc{tc} messages it originates.}

\hl{Two features follow directly. \mbox{\texttt{AverageAdvertisedLinksPerTCMessage}} is the mean number of links carried by a received \textsc{tc} message, with separately received flooded copies counted separately; a message originated under an active defense carries the injected link in addition to the real ones. \mbox{\texttt{AdvertisedLinksPerKnownNode}} normalizes the sum of each originator's most recent advertised-link count by the number of distinct node addresses the listener has seen; the injected address can contribute to both quantities.}

\hl{\mbox{\texttt{TcMessageRate}} responds indirectly. A node originates \textsc{tc} only while some neighbor has selected it as a multipoint relay, and a fictitious neighbor is reachable only through the node that advertises it, so that node is selected.}

\hl{Section~}\ref{subsec:protocol-footprint}\hl{ examines these expectations using paired raw control-plane measurements; \mbox{\texttt{AdvertisedLinksPerKnownNode}} is not measured separately there. All measured quantities move in the predicted direction. This is consistent with the proposed account but does not establish it causally, because advertised-link count and message volume were not varied independently.}

\hl{Taken together, the three accounts link the retained features to the defense's topology augmentation rather than to behavior of the Node Isolation Attack: the injected state enters the advertised links directly and affects \textsc{tc} origination through multipoint-relay selection.}

\subsection{\hl{Mobility and the Limits of the Separability Analyses}}
\label{subsec:mobility_impact}

\hl{Detection is more accurate under mobility for every tree-based, boosting and meta-ensemble model, and for the kernel-based \textsc{svm}, whereas the linear models are flat or lower} (Table~\ref{tab:in_domain_performance})\hl{. The separability analyses move the other way: the mean magnitude of per-feature Cohen's $d$ over \textsc{Universal-3} decreases} (Table~\ref{tab:per_feature_separability})\hl{, as do both joint measures, with $95\%$ intervals excluding zero} (Table~\ref{tab:joint_separability})\hl{.}

\hl{These observations are not in conflict because both joint measures summarize class separation along a single linear direction and are therefore most directly relevant to models confined to linear separation. Linear Discriminant Analysis, the classifier corresponding directly to the second measure, is essentially unchanged between configurations \mbox{($0.8746 \rightarrow 0.8784$)}, as is Ridge \mbox{($0.8751 \rightarrow 0.8791$)}, while Logistic Regression and both linear \textsc{svm}s decline. The evaluated nonlinear models all gain, including the kernel-based \textsc{svm} \mbox{($0.8843 \rightarrow 0.9085$)}. The gain therefore resides in structure not captured by the two joint statistics; the analyses performed here do not identify that structure.}

\hl{The decrease in the univariate summary is also concentrated in one observable: \mbox{\texttt{AdvertisedLinksPerKnownNode}} falls from \mbox{$1.1043$} to \mbox{$0.5538$}, whereas the other two change by less than \mbox{$0.05$}. It therefore does not indicate a general weakening of the retained observables under mobility.}

\subsection{\hl{What Makes an Observable Useful for Transfer}}
\label{subsec:transfer-determinants}

\hl{The analyses above rule out two simple explanations for what makes an observable useful for cross-configuration transfer. First, marginal stability does not predict transfer utility. \mbox{\texttt{AdvertisedLinksPerKnownNode}} changes more in univariate class separation than any other observable, yet all six importance criteria retain it in both configurations, and removing it alone substantially reduces transfer accuracy in all four classifier-direction combinations tested}
(Section~\ref{subsec:feature-ablation-and-augmentation})\hl{. Family-level ablations likewise show that the effect of removing observables depends on the classifier and transfer direction rather than simply on their stability. Cohen's $d$ summarizes an observable's marginal class separation, whereas the selection procedure ranks its contribution within a fitted multivariate model.}

\hl{Second, simply adding observables does not improve transfer. Above three observables, both transfer directions remain within a narrow band and \textsc{auc} is flat}
(Table~\ref{tab:optimal-k}, Figure~\ref{fig:optimal-k})\hl{, and source-configuration accuracy does not improve materially}
(Section~\ref{subsec:feature-size})\hl{. The additional observables therefore provide little incremental predictive value once \textsc{Universal-3} is present, although the sweep does not establish that they carry no discriminative information in isolation.}

\hl{The selection also does not require both configurations. Ranking on the source training partition alone returns the same three observables in both transfer directions, with zero-shot accuracies of \mbox{$0.8736$} and \mbox{$0.8563$}}
(Section~\ref{subsec:feature-consistency})\hl{. Agreement between the static and mobile rankings is therefore supporting evidence rather than a requirement of the selection procedure. The distinction matters for the threat model considered here, in which the adversary observes one deployment and not a matched pair. This control is limited to the evaluated configurations and is not claimed more widely.}

\hl{The Gauss-Markov comparison}
(Section~\ref{subsec:feature-size})\hl{ further limits this result: similar observables are largely recovered and in-domain detection remains strong, yet transfer becomes strongly asymmetric. Recovering a similar subset therefore does not imply symmetric transfer; the near-symmetry reported here is specific to the static and random-walk pair.}

\subsection{\hl{Scope, Limitations, and Generalizability}}
\label{subsec:mechanism_scope}

\hl{This work evaluates two defense mechanisms: the fictitious-node defense}~\cite{schweitzer2025achieving} \hl{against the Node Isolation Attack, and the independently proposed \textsc{fpnt} trust-based defense}~\cite{tan2015trust} \hl{against a black-hole attack. The second case shows that passive detectability is not confined to fictitious-node topology augmentation, but it is not a controlled comparison because both the defense and the attack differ.}

\hl{The dominant observable also differs between the two evaluations. \textsc{fpnt} attaches trust metadata to each \textsc{tc} message a node originates, enlarging the frames the observer hears; mean frame size is the most discriminative individual observable in that campaign. Overall detectability does not rest on these byte-volume observables, however: removing all four leaves accuracy above \mbox{$0.92$} in both configurations, and mean frame size is not selected for \textsc{gcop}/\textsc{gcohp}}
(Section~\ref{subsec:in-domain})\hl{. Different mechanisms therefore need not expose the same dominant signature.}

\hl{The sensitivity experiments of Section~}\ref{subsec:robustness}\hl{ bound rather than extend the result. We find no evidence that replacing deterministic range-based reception with log-distance path loss and Nakagami fading degrades detection when route viability is preserved, and enabling RTS/CTS changes accuracy little. Neither experiment supports robustness to deployment density.}

\hl{Because a topology-augmenting defense modifies the advertised topology through which it acts, other such defenses may also leave observable footprints, although their detectability remains an empirical question. More generally, persistent defense-induced control-plane modifications may be observable to a passive listener, but the result cannot be assumed to extend to other defenses, attacks, or network conditions.}

\hl{The evidence is limited to the two mechanisms and attacks evaluated here, a single passive vantage point, one routing protocol, $50$ nodes in a fixed deployment area, one physical-layer rate, and constant-bit-rate traffic. Other topology and traffic regimes may alter the available signal. Cryptographic authentication, reactive rate limiting, footprint-obscuring countermeasures, adaptive attackers, attack outcomes conditioned on prior detection, and defenses dormant until a threat is detected were not evaluated. The results are also limited to the connected \textsc{manet} regime represented by the experimental configuration. Generalization beyond these settings remains for future study.}

\section{Conclusion}
\label{sec:conclusion}

This study investigated whether deployed defense mechanisms can be identified from their impact on observable network traffic. The results demonstrate that defense detection is feasible in this setting. Ensemble classifiers reach in-domain accuracy of \hl{$0.8998$} under static conditions and \hl{$0.9298$} under mobility, with \textsc{auc} of \hl{$0.9403$ and $0.9729$} (Table~\ref{tab:in_domain_performance}). \hl{Without target-domain retraining or calibration, the three-observable subset reaches cross-domain accuracy of $0.8730$ from static to mobile and $0.8632$ from mobile to static} (Table~\ref{tab:optimal-k})\hl{. Together with the \textsc{fpnt} evaluation, these results show that the two defense mechanisms evaluated here, the fictitious-node defense under a node-isolation attack and the \textsc{fpnt} trust-based defense under a black-hole attack, can introduce changes that are detectable by a passive observer.}

\hl{Additional observables provide little incremental predictive value once the three-observable subset is present, and univariate stability does not identify which observables support transfer} (Table~\ref{tab:optimal-k}, Section~\ref{subsec:transfer-determinants})\hl{. The observable whose class separation changes most between configurations is retained by the multivariate selection and contributes to transfer performance.}

In the evaluated setting, routing-layer defenses of the type examined here, despite remaining protocol-compliant and externally unobtrusive, can leave detectable statistical traces in observable network behavior. Defense inference therefore functions as a reconnaissance capability: a passive adversary may use it to estimate whether \hl{one of these defenses} is active before committing to an attack strategy and adjust operational decisions accordingly.

Overall, \hl{the observables that transfer across the two configurations are those ranked highly by multivariate importance, which the source configuration alone already identifies, and they are control-plane quantities tied to the operation of the fictitious-node defense} (Section~\ref{subsec:mechanism})\hl{. A small set of protocol-level features observable from overheard transmissions supports cross-domain detection between the static and random-walk configurations evaluated here under the single-vantage passive observation model. Whether the same holds for other defense mechanisms, attacks, or mobility regimes was not tested and remains a question for future work.}

\section{Future Work}
\label{sec:future-work}

The present analysis selects \hl{a compact} cross-domain feature subset through a post-hoc \hl{procedure based on repeated run-grouped resampling and agreement between the two configurations}. A natural extension is to incorporate cross-domain \hl{robustness} directly into the feature-selection objective, for example by jointly optimizing in-domain discriminative power and cross-domain \hl{transfer performance}. \hl{Such an objective would have to be multivariate: the ablations reported in Section~}\ref{subsec:feature-ablation-and-augmentation}\hl{ show that the stability of a single observable's univariate class separation does not predict whether removing it harms transfer.} Such methods may recover \hl{transferable} subsets automatically across a broader range of operating conditions and reduce reliance on dataset-specific tuning.

\hl{A further question is raised by the mobility result itself. Detection improves under mobility while both joint separability measures decrease, and the analyses performed here do not identify the structure responsible} (Section~\ref{subsec:mobility_impact})\hl{. This could be investigated using separability measures not confined to a single linear direction and analyses of the nonlinear structure exploited by the classifiers.}

\hl{The present study evaluates two defense mechanisms, each against a different attack}~\cite{schweitzer2025achieving, tan2015trust}\hl{.} Extending the analysis to additional attack vectors and defense strategies would test whether the observed cross-domain feature patterns generalize beyond \hl{these settings, which the present evidence does not establish}. \hl{A direct test would hold the attack fixed and vary the defense, training a detector on one mechanism and evaluating it on another. This would help distinguish a footprint that transfers across mechanisms from one tied to a particular defense; the two evaluations reported here do not provide that design because the defense and attack change together.}

\hl{A further direction concerns the defender rather than the defense. The mechanisms evaluated here do not act to conceal themselves, and whether a concealment-aware defense could reduce its observable footprint without losing its protective effect was not examined. Evaluating it would require a defined adaptive-defender model together with the constraint that any concealment mechanism preserve the effectiveness of the underlying defense.}

The present evaluation is restricted to \textsc{olsr}, a proactive link-state protocol. \textsc{manet} deployments also use reactive protocols such as \textsc{aodv} and \textsc{dsr}, and hybrid designs such as \textsc{zrp}, whose control-plane dynamics differ substantially. Examining whether defense detection remains feasible under these protocol families, and whether a comparable \hl{compact} feature subset emerges, would clarify whether the present findings reflect \textsc{olsr}-specific routing dynamics or extend to other routing protocols.

Two formulation extensions follow from the binary detection setting considered here. The first is identifying the specific type of deployed defense, yielding a multi-class classification task. The second addresses concurrent deployment of multiple defenses, requiring a multi-label formulation. In such settings, observable routing behavior reflects the combined effect of several mechanisms, \hl{whose signatures may reinforce, mask, or partially cancel one another, making attribution} substantially more difficult than single-defense detection.

Finally, \hl{the propagation and medium-access sensitivity analyses of Section~}\ref{subsec:robustness}\hl{ bound rather than settle generalization across operating conditions. The evaluated configurations retain one traffic profile and one network density, and transfer across substantially different node counts or deployment areas was not tested. Future evaluation should therefore consider bursty or heterogeneous traffic, mobility models beyond the random-walk and Gauss-Markov settings evaluated here, such as reference-point group mobility, and interference and packet-loss regimes beyond those produced by the physical-layer settings examined here.}

\section*{Declaration of Generative AI and AI-Assisted Technologies}

During the preparation of this work, the authors used ChatGPT for language editing and manuscript refinement. The authors reviewed and edited all generated content as needed and take full responsibility for the content of the published article.

\bibliographystyle{elsarticle-num}
\bibliography{references}

@article{hassan2024advanced,
  title={Advanced intrusion detection in MANETs: A survey of machine learning and optimization techniques for mitigating black/gray hole attacks},
  author={Hassan, Saad M and Mohamad, Mohd Murtadha and Muchtar, Farkhana Binti},
  journal={IEEE Access},
  year={2024},
  publisher={IEEE}
}

@article{sharon2022tantra,
  title={Tantra: Timing-based adversarial network traffic reshaping attack},
  author={Sharon, Yam and Berend, David and Liu, Yang and Shabtai, Asaf and Elovici, Yuval},
  journal={IEEE Transactions on Information Forensics and Security},
  volume={17},
  pages={3225--3237},
  year={2022},
  publisher={IEEE}
}

@article{he2023adversarial,
  title={Adversarial machine learning for network intrusion detection systems: A comprehensive survey},
  author={He, Ke and Kim, Dan Dongseong and Asghar, Muhammad Rizwan},
  journal={IEEE Communications Surveys \& Tutorials},
  volume={25},
  number={1},
  pages={538--566},
  year={2023},
  publisher={IEEE}
}

@article{debicha2023tad,
  title={TAD: Transfer learning-based multi-adversarial detection of evasion attacks against network intrusion detection systems},
  author={Debicha, Islam and Bauwens, Richard and Debatty, Thibault and Dricot, Jean-Michel and Kenaza, Tayeb and Mees, Wim},
  journal={Future Generation Computer Systems},
  volume={138},
  pages={185--197},
  year={2023},
  publisher={Elsevier}
}

@article{cheng2011evasion,
  title={Evasion techniques: Sneaking through your intrusion detection/prevention systems},
  author={Cheng, Tsung-Huan and Lin, Ying-Dar and Lai, Yuan-Cheng and Lin, Po-Ching},
  journal={IEEE Communications Surveys \& Tutorials},
  volume={14},
  number={4},
  pages={1011--1020},
  year={2011},
  publisher={IEEE}
}

@misc{clausen2003rfc3626,
  title={RFC3626: Optimized link state routing protocol (OLSR)},
  author={Clausen, Thomas and Jacquet, Philippe},
  year={2003},
  publisher={RFC Editor}
}

@article{schweitzer2023persuasive,
  title={Persuasive: A node isolation attack variant for OLSR-based MANETs and its mitigation},
  author={Schweitzer, Nadav and Cohen, Liad and Dvir, Amit and Stulman, Ariel},
  journal={Ad Hoc Networks},
  volume={148},
  pages={103192},
  year={2023},
  publisher={Elsevier}
}

@article{bilot2023graph,
  title={Graph neural networks for intrusion detection: A survey},
  author={Bilot, Tristan and El Madhoun, Nour and Al Agha, Khaldoun and Zouaoui, Anis},
  journal={IEEE Access},
  volume={11},
  pages={49114--49139},
  year={2023},
  publisher={IEEE}
}

@article{saminathan2025multicast,
  title={Multicast On-Route cluster propagation to detect network intrusion detection systems on MANET using Deep Operator Neural networks},
  author={Saminathan, Karunakaran and Perumal, Latha and Shajin, Francis H and Shakya, Rajeev Kumar},
  journal={Expert Systems with Applications},
  volume={271},
  pages={125864},
  year={2025},
  publisher={Elsevier}
}

@article{zhu2021game,
  title={Game-theoretic and machine learning-based approaches for defensive deception: A survey},
  author={Zhu, Mu and Anwar, Ahmed H and Wan, Zelin and Cho, Jin-Hee and Kamhoua, Charles and Singh, Munindar P},
  journal={arXiv preprint arXiv:2101.10121},
  year={2021}
}

@article{zhang2021three,
  title={Three decades of deception techniques in active cyber defense-retrospect and outlook},
  author={Zhang, Li and Thing, Vrizlynn LL},
  journal={Computers \& Security},
  volume={106},
  pages={102288},
  year={2021},
  publisher={Elsevier}
}

@article{pawlick2019game,
  title={A game-theoretic taxonomy and survey of defensive deception for cybersecurity and privacy},
  author={Pawlick, Jeffrey and Colbert, Edward and Zhu, Quanyan},
  journal={ACM Computing Surveys (CSUR)},
  volume={52},
  number={4},
  pages={1--28},
  year={2019},
  publisher={ACM New York, NY, USA}
}

@article{hou2021combating,
  title={Combating adversarial network topology inference by proactive topology obfuscation},
  author={Hou, Tao and Wang, Tao and Lu, Zhuo and Liu, Yao},
  journal={IEEE/ACM Transactions on Networking},
  volume={29},
  number={6},
  pages={2779--2792},
  year={2021},
  publisher={IEEE}
}

@article{kong2022combination,
  title={Combination attacks and defenses on sdn topology discovery},
  author={Kong, Dezhang and Shen, Yi and Chen, Xiang and Cheng, Qiumei and Liu, Hongyan and Zhang, Dong and Liu, Xuan and Chen, Shuangxi and Wu, Chunming},
  journal={IEEE/ACM Transactions on Networking},
  volume={31},
  number={2},
  pages={904--919},
  year={2022},
  publisher={IEEE}
}

@article{shen2022machine,
  title={Machine learning-powered encrypted network traffic analysis: A comprehensive survey},
  author={Shen, Meng and Ye, Ke and Liu, Xingtong and Zhu, Liehuang and Kang, Jiawen and Yu, Shui and Li, Qi and Xu, Ke},
  journal={IEEE Communications Surveys \& Tutorials},
  volume={25},
  number={1},
  pages={791--824},
  year={2022},
  publisher={IEEE}
}

@inproceedings{kim2022equalnet,
  title={EqualNet: A Secure and Practical Defense for Long-term Network Topology Obfuscation.},
  author={Kim, Jinwoo and Marin, Eduard and Conti, Mauro and Shin, Seungwon},
  booktitle={NDSS},
  year={2022}
}

@article{li2025tunnel,
  title={Tunnel enabled programmable switches obfuscate network topology to defend against link flooding reconnaissance in software defined networking},
  author={Li, Xiang and Lee, Jungmin and Son, Junggab and Lee, Yeonjoon},
  journal={Scientific Reports},
  volume={15},
  number={1},
  pages={35549},
  year={2025},
  publisher={Nature Publishing Group UK London}
}

@article{schweitzer2025achieving,
  title={Achieving manet protection without the use of superfluous fictitious nodes},
  author={Schweitzer, Nadav and Cohen, Liad and Hirst, Tirza and Dvir, Amit and Stulman, Ariel},
  journal={Computer Communications},
  volume={229},
  pages={107978},
  year={2025},
  publisher={Elsevier}
}

@article{schweitzer2015mitigating,
  title={Mitigating denial of service attacks in OLSR protocol using fictitious nodes},
  author={Schweitzer, Nadav and Stulman, Ariel and Shabtai, Asaf and Margalit, Roy David},
  journal={IEEE Transactions on Mobile Computing},
  volume={15},
  number={1},
  pages={163--172},
  year={2015},
  publisher={IEEE}
}

@article{wang2023adversarial,
  title={Adversarial attacks and defenses in machine learning-empowered communication systems and networks: A contemporary survey},
  author={Wang, Yulong and Sun, Tong and Li, Shenghong and Yuan, Xin and Ni, Wei and Hossain, Ekram and Poor, H Vincent},
  journal={IEEE Communications Surveys \& Tutorials},
  volume={25},
  number={4},
  pages={2245--2298},
  year={2023},
  publisher={IEEE}
}

@article{tan2015trust,
  title={Trust based routing mechanism for securing OLSR-based MANET},
  author={Tan, Shuaishuai and Li, Xiaoping and Dong, Qingkuan},
  journal={Ad Hoc Networks},
  volume={30},
  pages={84--98},
  year={2015},
  publisher={Elsevier}
}

@article{tu2021active,
  title={An active-routing authentication scheme in MANET},
  author={Tu, Jinbin and Tian, Dahai and Wang, Yun},
  journal={IEEE Access},
  volume={9},
  pages={34276--34286},
  year={2021},
  publisher={IEEE}
}

@Misc{NS3,
  title        = {The ns-3 network simulator, version 3.47},
  howpublished = {\url{https://www.nsnam.org}},
  note         = {Accessed: 11 August 2026},
  year         = {2026},
}

@article{solorio2022survey,
  title={A survey on feature selection methods for mixed data},
  author={Solorio-Fern{\'a}ndez, Sa{\'u}l and Carrasco-Ochoa, J Ariel and Martinez-Trinidad, Jose Francisco},
  journal={Artificial Intelligence Review},
  volume={55},
  number={4},
  pages={2821--2846},
  year={2022},
  publisher={Springer}
}

@article{eltahlawy2023survey,
  title={A survey on parameters affecting MANET performance},
  author={Eltahlawy, Ahmed M and Aslan, Heba K and Abdallah, Eslam G and Elsayed, Mahmoud Said and Jurcut, Anca D and Azer, Marianne A},
  journal={Electronics},
  volume={12},
  number={9},
  pages={1956},
  year={2023},
  publisher={MDPI}
}

@article{quy2019survey,
  title={Survey of Recent Routing Metrics and Protocols for Mobile Ad-Hoc Networks.},
  author={Quy, Vu Khanh and Ban, Nguyen Tien and Nam, Vi Hoai and Tuan, Dao Minh and Han, Nguyen Dinh},
  journal={J. Commun.},
  volume={14},
  number={2},
  pages={110--120},
  year={2019}
}

@article{jeniffer2025efficient,
  title={Efficient Intrusion Detection in Wireless Sensor Networks Using MH-CEGRU with Cross-Layer Monitoring and Cryptographic Security},
  author={Jeniffer, J Thresa and Chandrasekar, A and Radhika, S},
  journal={Knowledge-Based Systems},
  pages={114707},
  year={2025},
  publisher={Elsevier}
}

@article{zohourian2024iot,
  title={IoT-PRIDS: Leveraging packet representations for intrusion detection in IoT networks},
  author={Zohourian, Alireza and Dadkhah, Sajjad and Molyneaux, Heather and Neto, Euclides Carlos Pinto and Ghorbani, Ali A},
  journal={Computers \& Security},
  volume={146},
  pages={104034},
  year={2024},
  publisher={Elsevier}
}

@inproceedings{shen2024effective,
  title={Effective intrusion detection in heterogeneous Internet-of-Things networks via ensemble knowledge distillation-based federated learning},
  author={Shen, Jiyuan and Yang, Wenzhuo and Chu, Zhaowei and Fan, Jiani and Niyato, Dusit and Lam, Kwok-Yan},
  booktitle={ICC 2024-IEEE International Conference on Communications},
  pages={2034--2039},
  year={2024},
  organization={IEEE}
}

@article{sefati2025comprehensive,
  title={A comprehensive survey of cybersecurity techniques based on quality of service (QoS) on the Internet of Things (IoT)},
  author={Sefati, Seyed Salar and Arasteh, Bahman and Halunga, Simona and Fratu, Octavian},
  journal={Cluster Computing},
  volume={28},
  number={12},
  pages={792},
  year={2025},
  publisher={Springer}
}

@article{wang2025survey,
  title={A survey on Identity and Access Management for future IoT services},
  author={Wang, Yiting and Castillejo, Pedro and Mart{\'\i}nez-Ortega, Jos{\'e}-Fern{\'a}n and D{\'\i}az, Vicente Hern{\'a}ndez},
  journal={Computer Networks},
  pages={111718},
  year={2025},
  publisher={Elsevier}
}

@article{singh2022cryptographic,
  title={A cryptographic approach to prevent network incursion for enhancement of QoS in sustainable smart city using MANET},
  author={Singh, Saurabh and Pise, Anil and Alfarraj, Osama and Tolba, Amr and Yoon, Byungun},
  journal={Sustainable Cities and Society},
  volume={79},
  pages={103483},
  year={2022},
  publisher={Elsevier}
}

@article{tabatabaei2023introducing,
  title={Introducing a new routing method in the MANET using the symbionts search algorithm},
  author={Tabatabaei, Shayesteh},
  journal={Plos one},
  volume={18},
  number={8},
  pages={e0290091},
  year={2023},
  publisher={Public Library of Science San Francisco, CA USA}
}

@article{kim2023extended,
  title={Extended data plane architecture for in-network security services in software-defined networks},
  author={Kim, Jinwoo and Kim, Yeonkeun and Yegneswaran, Vinod and Porras, Phillip and Shin, Seungwon and Park, Taejune},
  journal={Computers \& Security},
  volume={124},
  pages={102976},
  year={2023},
  publisher={Elsevier}
}

@article{afraji2025deep,
  title={Deep learning-driven defense strategies for mitigating DDoS attacks in cloud computing environments},
  author={Afraji, Doaa Mohsin Abd Ali and Lloret, Jaime and Pe{\~n}alver, Lourdes},
  journal={Cyber Security and Applications},
  volume={3},
  pages={100085},
  year={2025},
  publisher={Elsevier}
}

@article{czaja2025cybersecurity,
  title={Cybersecurity challenges and opportunities of machine learning-based artificial intelligence},
  author={Czaja, Pawel and Gdowski, Bartlomiej and Niemiec, Marcin and Mees, Wim and Stoianov, Nikolai and Votis, Konstantinos and Kharchenko, Vyacheslav and Katos, Vasilis and Merialdo, Matteo},
  journal={Neural Computing and Applications},
  volume={37},
  number={33},
  pages={27931--27956},
  year={2025},
  publisher={Springer}
}

@article{tebbaa2026mitigating,
  title={Mitigating DDoS attacks in software-defined networks: a systematic literature review of machine learning and deep learning approaches},
  author={Tebbaa, Kaoutar and Chakir, Oumaima and Maleh, Yassine and Belaissaoui, Mustapha},
  journal={Iran Journal of Computer Science},
  volume={9},
  number={1},
  pages={5},
  year={2026},
  publisher={Springer}
}

@article{shwartz2022tabular,
  title={Tabular data: Deep learning is not all you need},
  author={Shwartz-Ziv, Ravid and Armon, Amitai},
  journal={Information Fusion},
  volume={81},
  pages={84--90},
  year={2022}
}

@inproceedings{grinsztajn2022why,
  title={Why do tree-based models still outperform deep learning on typical tabular data?},
  author={Grinsztajn, L{\'e}o and Oyallon, Edouard and Varoquaux, Ga{\"e}l},
  booktitle={NeurIPS},
  year={2022}
}

@article{probst2019tunability,
  title={Tunability: Importance of Hyperparameters of Machine Learning Algorithms},
  author={Probst, Philipp and Boulesteix, Anne-Laure and Bischl, Bernd},
  journal={Journal of Machine Learning Research},
  volume={20},
  number={53},
  pages={1--32},
  year={2019}
}

@inproceedings{zadrozny2002transforming,
  title={Transforming classifier scores into accurate multiclass probability estimates},
  author={Zadrozny, Bianca and Elkan, Charles},
  booktitle={KDD},
  pages={694--699},
  year={2002}
}

@inproceedings{niculescu2005predicting,
  title={Predicting good probabilities with supervised learning},
  author={Niculescu-Mizil, Alexandru and Caruana, Rich},
  booktitle={ICML},
  pages={625--632},
  year={2005}
}

@article{wolpert1992stacked,
  title={Stacked generalization},
  author={Wolpert, David H.},
  journal={Neural Networks},
  volume={5},
  number={2},
  pages={241--259},
  year={1992}
}

@ARTICLE{saedi2026zerotrust,
  author={Saedi, Mohammad and Moore, Adrian and Jaafar, Fehmi and Rahman, Md. Arafatur and Biswas, Sujit},
  journal={IEEE Open Journal of the Communications Society}, 
  title={Zero Trust Rule-Based Protocol for Secure 5G Handover Against Rogue Base Stations}, 
  year={2026},
  volume={7},
  number={},
  pages={5092-5106},
  doi={10.1109/OJCOMS.2026.3691214}
}

@ARTICLE{mpitziopoulos2009jamming,
  author={Mpitziopoulos, Aristides and Gavalas, Damianos and Konstantopoulos, Charalampos and Pantziou, Grammati},
  journal={IEEE Communications Surveys \& Tutorials}, 
  title={A survey on jamming attacks and countermeasures in WSNs}, 
  year={2009},
  volume={11},
  number={4},
  pages={42-56},
  doi={10.1109/SURV.2009.090404}
}

@Article{saedi2022synthetic,
  AUTHOR = {Saedi, Mohammad and Moore, Adrian and Perry, Philip},
  title = {Synthetic Generation of Realistic Signal Strength Data to Enable 5G Rogue Base Station Investigation in Vehicular Platooning},
  JOURNAL = {Applied Sciences},
  VOLUME = {12},
  YEAR = {2022},
  NUMBER = {24},
  ARTICLE-NUMBER = {12516},
  URL = {https://www.mdpi.com/2076-3417/12/24/12516},
  ISSN = {2076-3417},
  DOI = {10.3390/app122412516}
}

\appendix
\section{Hyperparameter Details}

\begin{table*}[t]
\centering
\footnotesize
\caption{Hyperparameter search spaces used by the unified procedure of Section~\ref{subsec:classifiers-hyperparameters} (Part I: tree-based and boosting models). Tree-based, boosting, and bagging models were tuned using \texttt{RandomizedSearchCV} ($n_{\mathrm{iter}}=80$). All searches used 3-fold stratified cross-validation with \texttt{random\_state}=42. Values $\infty$ and $-1$ denote unbounded depth.}
\label{tab:search-space-a}

\begin{tabular}{lll}
\toprule
Model & Hyperparameter & Search range \\
\midrule

\multirow{5}{*}{Random Forest}
& \texttt{n\_estimators} & $\{300,500,800,1200,1600\}$ \\
& \texttt{max\_depth} & $\{10,15,20,30,\infty\}$ \\
& \texttt{min\_samples\_split} & $\{2,5,10,20\}$ \\
& \texttt{min\_samples\_leaf} & $\{1,2,5,10\}$ \\
& \texttt{max\_features} & $\{\text{sqrt},\text{log2},0.3,0.5,0.7\}$ \\
\midrule

\multirow{5}{*}{Extra Trees}
& \texttt{n\_estimators} & $\{300,500,800,1200,1600\}$ \\
& \texttt{max\_depth} & $\{10,15,20,30,\infty\}$ \\
& \texttt{min\_samples\_split} & $\{2,5,10,20\}$ \\
& \texttt{min\_samples\_leaf} & $\{1,2,5,10\}$ \\
& \texttt{max\_features} & $\{\text{sqrt},\text{log2},0.3,0.5,0.7\}$ \\
\midrule

\multirow{5}{*}{Gradient Boosting}
& \texttt{n\_estimators} & $\{300,500,800,1200\}$ \\
& \texttt{learning\_rate} & $\{0.01,0.03,0.05,0.1,0.2\}$ \\
& \texttt{max\_depth} & $\{3,4,6,7,9,11\}$ \\
& \texttt{subsample} & $\{0.5,0.7,0.8,1.0\}$ \\
& \texttt{max\_features} & $\{\text{sqrt},\text{log2},0.3,0.5,0.7\}$ \\
\midrule

\multirow{7}{*}{\textsc{xgboost}}
& \texttt{n\_estimators} & $\{500,800,1000,1500,2000\}$ \\
& \texttt{learning\_rate} & $\{0.01,0.03,0.05,0.1,0.2\}$ \\
& \texttt{max\_depth} & $\{3,4,6,7,9,11\}$ \\
& \texttt{subsample} & $\{0.5,0.7,0.8,1.0\}$ \\
& \texttt{colsample\_bytree} & $\{0.5,0.7,0.8,1.0\}$ \\
& \texttt{min\_child\_weight} & $\{1,3,5,10\}$ \\
& \texttt{reg\_lambda} & $\{0.1,1,3,5,10\}$ \\
\midrule

\multirow{5}{*}{\textsc{catboost}}
& \texttt{iterations} & $\{500,800,1000,1500\}$ \\
& \texttt{learning\_rate} & $\{0.01,0.03,0.05,0.1,0.2\}$ \\
& \texttt{depth} & $\{4,6,7,8,9,10\}$ \\
& \texttt{l2\_leaf\_reg} & $\{0.5,1,3,5,10,20\}$ \\
& \texttt{subsample} & $\{0.5,0.7,0.8,1.0\}$ \\
\midrule

\multirow{8}{*}{\textsc{lightgbm}}
& \texttt{n\_estimators} & $\{500,800,1000,1500,2000\}$ \\
& \texttt{learning\_rate} & $\{0.01,0.03,0.05,0.1,0.2\}$ \\
& \texttt{max\_depth} & $\{4,6,8,10,12,-1\}$ \\
& \texttt{num\_leaves} & $\{15,31,63,127,255\}$ \\
& \texttt{subsample} & $\{0.5,0.7,0.8,1.0\}$ \\
& \texttt{colsample\_bytree} & $\{0.5,0.7,0.8,1.0\}$ \\
& \texttt{min\_child\_samples} & $\{5,10,20,50,100\}$ \\
& \texttt{reg\_lambda} & $\{0.0,0.1,0.5,1.0,5.0\}$ \\
\bottomrule
\end{tabular}
\end{table*}

\begin{table*}[t]
\centering
\footnotesize
\caption{Hyperparameter search spaces used by the unified procedure of Section~\ref{subsec:classifiers-hyperparameters} (Part II: meta-estimators and linear models). Logistic Regression, \textsc{svm}, and Ridge were tuned using \texttt{GridSearchCV} (full coverage).}
\label{tab:search-space-b}

\begin{tabular}{lll}
\toprule
Model & Hyperparameter & Search range \\
\midrule

\multirow{3}{*}{AdaBoost}
& \texttt{n\_estimators} & $\{50,100,200,300,500\}$ \\
& \texttt{learning\_rate} & $\{0.01,0.05,0.1,0.5,1.0,1.5\}$ \\
& base tree \texttt{max\_depth} & $\{1,3,5,7,10\}$ \\
\midrule

\multirow{4}{*}{Bagging-RF}
& \texttt{n\_estimators} & $\{50,100,200,300,500\}$ \\
& \texttt{max\_samples} & $\{0.3,0.5,0.7,0.8,1.0\}$ \\
& \texttt{max\_features} & $\{0.3,0.5,0.7,0.8,1.0\}$ \\
& base tree \texttt{max\_depth} & $\{10,15,20,30,\infty\}$ \\
\midrule

\multirow{4}{*}{Bagging-ET}
& \texttt{n\_estimators} & $\{20,30,50,100,200\}$ \\
& \texttt{max\_samples} & $\{0.3,0.5,0.7,0.8,1.0\}$ \\
& inner-tree \texttt{n\_estimators} & $\{10,20,30,50\}$ \\
& inner-tree \texttt{max\_depth} & $\{10,15,20,\infty\}$ \\
\midrule

\multirow{3}{*}{Logistic Regression}
& \texttt{C} & $\{10^{-4},10^{-3},10^{-2},10^{-1},1,10,10^{2},10^{3},10^{4}\}$ \\
& \texttt{penalty} & $\{\mathrm{L2}\}$ \\
& \texttt{solver} & $\{\mathrm{lbfgs},\mathrm{liblinear}\}$ \\
\midrule
\multirow{2}{*}{\textsc{svm} (\textsc{rbf})}
& \texttt{C} & $\{10^{-1},1,10,10^{2},10^{3},10^{4},10^{5}\}$ \\
& \texttt{gamma} & $\{10^{-7},10^{-6},10^{-5},10^{-4},10^{-3},10^{-2},0.1,1,\mathrm{scale}\}$ \\
\midrule
Ridge
& \texttt{alpha} & $\{10^{-4},10^{-3},10^{-2},10^{-1},1,10,10^{2},10^{3},10^{4}\}$ \\
\bottomrule
\end{tabular}
\end{table*}

\begin{table*}[t]
\centering
\footnotesize
\caption{\hl{Hyperparameter configurations selected by the unified search of Section~}\ref{subsec:classifiers-hyperparameters}\hl{, for the static setting.}}
\label{tab:hyperparameters_static}
\begin{tabular}{ll}
\toprule
Model                       & Static configuration \\
\midrule
Random Forest               & 800 trees, max depth 30, min split 10, min leaf 10, max features 0.5 \\
Extra Trees                 & 300 trees, max depth 30, min split 10, min leaf 1, max features 0.5 \\
Gradient Boosting           & 1{,}200 estimators, learning rate 0.03, max depth 4, subsample 1.0 \\
\textsc{xgboost}            & 2{,}000 rounds, learning rate 0.01, max depth 7, subsample 0.7 \\
\textsc{catboost}           & 500 iterations, learning rate 0.05, depth 7 \\
\textsc{lightgbm}           & 800 estimators, learning rate 0.01, num leaves 63 \\
AdaBoost                    & 500 estimators, learning rate 0.1 \\
Bagging-RF                  & 100 base estimators, max depth 15 \\
Bagging-ET                  & 20 base estimators, max depth $\infty$ \\
Logistic Regression         & L2 penalty, $C = 10$ \\
\textsc{svm} (\textsc{rbf}) & \hl{$C = 10^{4}$, $\gamma = 10^{-3}$} \\
Ridge                       & $\alpha = 0.01$ \\
Stacking Ensemble           & 10 base learners, Logistic Regression meta-learner \\
\bottomrule
\end{tabular}
\end{table*}

\begin{table*}[t]
\centering
\footnotesize
\caption{\hl{Hyperparameter configurations selected by the unified search of Section~}\ref{subsec:classifiers-hyperparameters}\hl{, for the mobile setting.}}
\label{tab:hyperparameters_mobile}
\begin{tabular}{ll}
\toprule
Model                       & Mobile configuration \\
\midrule
Random Forest               & 800 trees, max depth $\infty$, min split 5, min leaf 1, max features 0.7 \\
Extra Trees                 & 800 trees, max depth $\infty$, min split 5, min leaf 1, max features 0.7 \\
Gradient Boosting           & 1{,}200 estimators, learning rate 0.05, max depth 4, subsample 1.0 \\
\textsc{xgboost}            & 2{,}000 rounds, learning rate 0.05, max depth 3, subsample 0.8 \\
\textsc{catboost}           & 1{,}000 iterations, learning rate 0.05, depth 4 \\
\textsc{lightgbm}           & 500 estimators, learning rate 0.05, num leaves 63 \\
AdaBoost                    & 300 estimators, learning rate 0.5 \\
Bagging-RF                  & 200 base estimators, max depth 30 \\
Bagging-ET                  & 30 base estimators, max depth $\infty$ \\
Logistic Regression         & L2 penalty, $C = 10^{-4}$ \\
\textsc{svm} (\textsc{rbf}) & \hl{$C = 10^{2}$, $\gamma = 10^{-3}$} \\
Ridge                       & $\alpha = 1.0$ \\
Stacking Ensemble & same as static \\
\bottomrule
\end{tabular}
\end{table*}

\clearpage
\section{Cost-Sensitive Threshold Analysis}

\begin{table}[!ht]
\centering
\footnotesize
\setlength{\tabcolsep}{4pt}
\caption{Cost-sensitive operating points for the Stacking Ensemble. Expected cost per measurement window is $(\mathrm{FP} + C\cdot\mathrm{FN})/N$, where a false alarm costs $1$ and a missed defense costs $C$; $C=1$ reduces to the error rate. For each $C$ the cost-minimizing threshold is selected on the threshold-selection half of the validation partition, locked, and only then evaluated on the test partition; no test label is consulted during selection and nothing is retrained. \emph{Saving} is the reduction in expected cost relative to the reported operating point, the accuracy-selected threshold of $0.640$ (static) and $0.480$ (mobile), whose cost is $(75 + 727\,C)/8000$ and $(161 + 401\,C)/8000$ respectively. Values after $\pm$ are standard deviations of the paired saving over $2{,}000$ cluster resamples at the simulation-run level; $n = 2{,}000$ test runs and $8{,}000$ measurement windows per configuration.}
\label{tab:cost-sensitive}
\begin{tabular}{lcccccccc}
\toprule
 & \multicolumn{4}{c}{Static} & \multicolumn{4}{c}{Mobile} \\
\cmidrule(lr){2-5}\cmidrule(lr){6-9}
$C$ & Thr. & Cost & Acc. & Saving & Thr. & Cost & Acc. & Saving \\
\midrule
0.25 & 0.78 & 0.0297 & 0.8970 & $+0.0024 \pm 0.0008$ & 0.93 & 0.0255 & 0.9146 & $+0.0072 \pm 0.0014$ \\
0.5  & 0.78 & 0.0541 & 0.8970 & $+0.0007 \pm 0.0009$ & 0.76 & 0.0439 & 0.9241 & $+0.0013 \pm 0.0012$ \\
1    & 0.64 & 0.1003 & 0.8998 & $\phantom{+}0.0000 \pm 0.0000$ & 0.48 & 0.0703 & 0.9298 & $\phantom{+}0.0000 \pm 0.0000$ \\
2    & 0.30 & 0.1777 & 0.8930 & $+0.0134 \pm 0.0037$ & 0.17 & 0.1156 & 0.9153 & $+0.0047 \pm 0.0038$ \\
4    & 0.16 & 0.3031 & 0.8431 & $+0.0698 \pm 0.0106$ & 0.13 & 0.1770 & 0.8999 & $+0.0436 \pm 0.0076$ \\
10   & 0.08 & 0.4753 & 0.6541 & $+0.4429 \pm 0.0364$ & 0.08 & 0.2995 & 0.8107 & $+0.2219 \pm 0.0234$ \\
\bottomrule
\end{tabular}
\end{table}

\clearpage
\section{\hl{The Engineered Feature Set}}

\hl{Table~}\ref{tab:feature-list}\hl{ enumerates the expanded representation of Section~}\ref{subsec:metric-collection}\hl{. Every column beyond the seventeen observables is derived from those seventeen alone, so none depends on a quantity outside the observable space. Several definitions state what the quantity is not, because the column keeps the name of a network-wide metric that a single vantage point cannot compute.}

{\footnotesize
\setlength{\tabcolsep}{4pt}
\begin{longtable}{@{}r>{\ttfamily\raggedright\arraybackslash}p{0.30\linewidth}>{\raggedright\arraybackslash}p{0.40\linewidth}cc@{}}
\caption{\hl{The 67 engineered columns, with the configuration that retains each. The seventeen single-vantage observables are listed first; the remaining fifty are derived from them alone. Names are those used throughout this paper.}}
\label{tab:feature-list}\\
\toprule
\# & Feature & Definition & Static & Mobile \\
\midrule
\endfirsthead
\toprule
\# & Feature & Definition & Static & Mobile \\
\midrule
\endhead
\bottomrule
\endfoot
\addlinespace[3pt]
\multicolumn{5}{@{}l@{}}{\textbf{The seventeen single-vantage observables} (17)} \\
\addlinespace[4pt]
1 & Tc\allowbreak{}Message\allowbreak{}Rate & Distinct TC messages received by the observer per second, de-duplicated across flooded copies by (originator, sequence number). & \checkmark & \checkmark \\
2 & Average\allowbreak{}Advertised\allowbreak{}Links\allowbreak{}Per\allowbreak{}TCMessage & Mean number of links advertised per TC message heard. & \checkmark & \checkmark \\
3 & Mid\allowbreak{}Message\allowbreak{}Rate & MID messages heard per second. &  &  \\
4 & Hna\allowbreak{}Message\allowbreak{}Rate & HNA messages heard per second. &  &  \\
5 & Routing\allowbreak{}Overhead\allowbreak{}Bytes\allowbreak{}Ratio & OLSR control bytes divided by data bytes, over the frames heard. & \checkmark & \checkmark \\
6 & Advertised\allowbreak{}Links\allowbreak{}Per\allowbreak{}Known\allowbreak{}Node & Sum of advertised links, over the number of distinct node addresses this vantage point has seen. NOT an MPR-selector count: under an active defence the advertised set carries the injected fictitious link, which the node's real selector set does not. & \checkmark & \checkmark \\
7 & Average\allowbreak{}Hop\allowbreak{}Count & Mean of the hop-count field in the headers of the TC messages received, i.e. how far in hops the TC originators heard were. NOT the mean data-path hop count. & \checkmark & \checkmark \\
8 & Sniffed\allowbreak{}Frame\allowbreak{}Rate & All frames successfully received at the PHY per second, including acknowledgements and control frames. & \checkmark & \checkmark \\
9 & Observed\allowbreak{}Data\allowbreak{}Rate & Bit rate of the IPv4-carrying frames heard at this vantage point. Not a network-wide throughput. & \checkmark & \checkmark \\
10 & Control\allowbreak{}To\allowbreak{}Data\allowbreak{}Frame\allowbreak{}Ratio & OLSR frames per non-OLSR data frame heard. The conventional normalised routing load divides by packets delivered, which a listener cannot observe. & \checkmark & \checkmark \\
11 & Mean\allowbreak{}Sniffed\allowbreak{}Frame\allowbreak{}Size & Mean size of a frame received at this vantage point. A listener observes each frame once on the medium and cannot separate transmitted from received copies. & \checkmark & \checkmark \\
12 & Mean\allowbreak{}Per\allowbreak{}Source\allowbreak{}Data\allowbreak{}Rate & Mean data rate over the sources this vantage point overheard. Per source, not per classified flow. & \checkmark & \checkmark \\
13 & Per\allowbreak{}Source\allowbreak{}Rate\allowbreak{}Std & Standard deviation of the per-source byte rates over the sources heard. The network-wide metric of the same name is computed by FlowMonitor over 51 classified flows from receiver-side rxBytes. & \checkmark & \checkmark \\
14 & Mean\allowbreak{}Source\allowbreak{}Visibility\allowbreak{}Window & Mean of (last heard - first heard) per source. A visibility window, not a flow lifetime. & \checkmark & \checkmark \\
15 & Source\allowbreak{}Visibility\allowbreak{}Window\allowbreak{}Std & Standard deviation of the per-source visibility windows. & \checkmark & \checkmark \\
16 & Mean\allowbreak{}Bytes\allowbreak{}Per\allowbreak{}Source & Mean bytes heard per source. Per source, not per flow; Tx and Rx are not separable at one vantage point. &  &  \\
17 & Mean\allowbreak{}Frames\allowbreak{}Per\allowbreak{}Source & Mean frames heard per source. & \checkmark & \checkmark \\
\addlinespace[6pt]
\midrule
\addlinespace[3pt]
\multicolumn{5}{@{}l@{}}{\textbf{Products and sums of the observables} (2)} \\
\addlinespace[4pt]
18 & TDR & how much topology each second carries: message rate times links per message & \checkmark & \checkmark \\
19 & Total\_\allowbreak{}Traffic & every message rate the listener hears, added together & \checkmark & \checkmark \\
\addlinespace[6pt]
\midrule
\addlinespace[3pt]
\multicolumn{5}{@{}l@{}}{\textbf{Log1p of each observable} (17)} \\
\addlinespace[4pt]
20 & log1p\_\allowbreak{}Tc\allowbreak{}Message\allowbreak{}Rate & log1p of TcMessageRate &  &  \\
21 & log1p\_\allowbreak{}Average\allowbreak{}Advertised\allowbreak{}Links\allowbreak{}Per\allowbreak{}TCMessage & log1p of AverageAdvertisedLinksPerTCMessage &  &  \\
22 & log1p\_\allowbreak{}Mid\allowbreak{}Message\allowbreak{}Rate & log1p of MidMessageRate &  &  \\
23 & log1p\_\allowbreak{}Hna\allowbreak{}Message\allowbreak{}Rate & log1p of HnaMessageRate &  &  \\
24 & log1p\_\allowbreak{}Routing\allowbreak{}Overhead\allowbreak{}Bytes\allowbreak{}Ratio & log1p of RoutingOverheadBytesRatio &  &  \\
25 & log1p\_\allowbreak{}Advertised\allowbreak{}Links\allowbreak{}Per\allowbreak{}Known\allowbreak{}Node & log1p of AdvertisedLinksPerKnownNode &  &  \\
26 & log1p\_\allowbreak{}Average\allowbreak{}Hop\allowbreak{}Count & log1p of AverageHopCount &  &  \\
27 & log1p\_\allowbreak{}Sniffed\allowbreak{}Frame\allowbreak{}Rate & log1p of SniffedFrameRate &  &  \\
28 & log1p\_\allowbreak{}Observed\allowbreak{}Data\allowbreak{}Rate & log1p of ObservedDataRate & \checkmark & \checkmark \\
29 & log1p\_\allowbreak{}Control\allowbreak{}To\allowbreak{}Data\allowbreak{}Frame\allowbreak{}Ratio & log1p of ControlToDataFrameRatio & \checkmark & \checkmark \\
30 & log1p\_\allowbreak{}Mean\allowbreak{}Sniffed\allowbreak{}Frame\allowbreak{}Size & log1p of MeanSniffedFrameSize &  & \checkmark \\
31 & log1p\_\allowbreak{}Mean\allowbreak{}Per\allowbreak{}Source\allowbreak{}Data\allowbreak{}Rate & log1p of MeanPerSourceDataRate &  & \checkmark \\
32 & log1p\_\allowbreak{}Per\allowbreak{}Source\allowbreak{}Rate\allowbreak{}Std & log1p of PerSourceRateStd & \checkmark & \checkmark \\
33 & log1p\_\allowbreak{}Mean\allowbreak{}Source\allowbreak{}Visibility\allowbreak{}Window & log1p of MeanSourceVisibilityWindow &  & \checkmark \\
34 & log1p\_\allowbreak{}Source\allowbreak{}Visibility\allowbreak{}Window\allowbreak{}Std & log1p of SourceVisibilityWindowStd &  &  \\
35 & log1p\_\allowbreak{}Mean\allowbreak{}Bytes\allowbreak{}Per\allowbreak{}Source & log1p of MeanBytesPerSource &  &  \\
36 & log1p\_\allowbreak{}Mean\allowbreak{}Frames\allowbreak{}Per\allowbreak{}Source & log1p of MeanFramesPerSource &  & \checkmark \\
\addlinespace[6pt]
\midrule
\addlinespace[3pt]
\multicolumn{5}{@{}l@{}}{\textbf{Square root of each observable} (17)} \\
\addlinespace[4pt]
37 & sqrt\_\allowbreak{}Tc\allowbreak{}Message\allowbreak{}Rate & square root of TcMessageRate &  &  \\
38 & sqrt\_\allowbreak{}Average\allowbreak{}Advertised\allowbreak{}Links\allowbreak{}Per\allowbreak{}TCMessage & square root of AverageAdvertisedLinksPerTCMessage &  &  \\
39 & sqrt\_\allowbreak{}Mid\allowbreak{}Message\allowbreak{}Rate & square root of MidMessageRate &  &  \\
40 & sqrt\_\allowbreak{}Hna\allowbreak{}Message\allowbreak{}Rate & square root of HnaMessageRate &  &  \\
41 & sqrt\_\allowbreak{}Routing\allowbreak{}Overhead\allowbreak{}Bytes\allowbreak{}Ratio & square root of RoutingOverheadBytesRatio &  &  \\
42 & sqrt\_\allowbreak{}Advertised\allowbreak{}Links\allowbreak{}Per\allowbreak{}Known\allowbreak{}Node & square root of AdvertisedLinksPerKnownNode &  &  \\
43 & sqrt\_\allowbreak{}Average\allowbreak{}Hop\allowbreak{}Count & square root of AverageHopCount &  &  \\
44 & sqrt\_\allowbreak{}Sniffed\allowbreak{}Frame\allowbreak{}Rate & square root of SniffedFrameRate &  &  \\
45 & sqrt\_\allowbreak{}Observed\allowbreak{}Data\allowbreak{}Rate & square root of ObservedDataRate &  &  \\
46 & sqrt\_\allowbreak{}Control\allowbreak{}To\allowbreak{}Data\allowbreak{}Frame\allowbreak{}Ratio & square root of ControlToDataFrameRatio &  & \checkmark \\
47 & sqrt\_\allowbreak{}Mean\allowbreak{}Sniffed\allowbreak{}Frame\allowbreak{}Size & square root of MeanSniffedFrameSize &  &  \\
48 & sqrt\_\allowbreak{}Mean\allowbreak{}Per\allowbreak{}Source\allowbreak{}Data\allowbreak{}Rate & square root of MeanPerSourceDataRate &  &  \\
49 & sqrt\_\allowbreak{}Per\allowbreak{}Source\allowbreak{}Rate\allowbreak{}Std & square root of PerSourceRateStd & \checkmark & \checkmark \\
50 & sqrt\_\allowbreak{}Mean\allowbreak{}Source\allowbreak{}Visibility\allowbreak{}Window & square root of MeanSourceVisibilityWindow &  &  \\
51 & sqrt\_\allowbreak{}Source\allowbreak{}Visibility\allowbreak{}Window\allowbreak{}Std & square root of SourceVisibilityWindowStd &  &  \\
52 & sqrt\_\allowbreak{}Mean\allowbreak{}Bytes\allowbreak{}Per\allowbreak{}Source & square root of MeanBytesPerSource &  &  \\
53 & sqrt\_\allowbreak{}Mean\allowbreak{}Frames\allowbreak{}Per\allowbreak{}Source & square root of MeanFramesPerSource &  &  \\
\addlinespace[6pt]
\midrule
\addlinespace[3pt]
\multicolumn{5}{@{}l@{}}{\textbf{Powers of the observed data rate} (2)} \\
\addlinespace[4pt]
54 & Observed\allowbreak{}Data\allowbreak{}Rate\_\allowbreak{}squared & ObservedDataRate, squared & \checkmark & \checkmark \\
55 & Observed\allowbreak{}Data\allowbreak{}Rate\_\allowbreak{}cubed & ObservedDataRate, cubed & \checkmark & \checkmark \\
\addlinespace[6pt]
\midrule
\addlinespace[3pt]
\multicolumn{5}{@{}l@{}}{\textbf{Statistics across the observables of one window} (12)} \\
\addlinespace[4pt]
56 & row\_\allowbreak{}mean & mean of the seventeen observables in this window & \checkmark &  \\
57 & row\_\allowbreak{}std & spread of the seventeen, as a standard deviation &  &  \\
58 & row\_\allowbreak{}median & middle value of the seventeen & \checkmark & \checkmark \\
59 & row\_\allowbreak{}max & largest of the seventeen &  &  \\
60 & row\_\allowbreak{}min & smallest of the seventeen &  &  \\
61 & row\_\allowbreak{}range & largest minus smallest &  &  \\
62 & row\_\allowbreak{}cv & spread relative to the mean, row\_std / row\_mean & \checkmark & \checkmark \\
63 & row\_\allowbreak{}skew & how lopsided the seventeen are about their mean & \checkmark & \checkmark \\
64 & row\_\allowbreak{}kurtosis & how heavy the tails of the seventeen are &  &  \\
65 & row\_\allowbreak{}q25 & value below which a quarter of the seventeen fall & \checkmark & \checkmark \\
66 & row\_\allowbreak{}q75 & value below which three quarters fall & \checkmark & \checkmark \\
67 & row\_\allowbreak{}iqr & row\_q75 minus row\_q25, the middle half's width &  &  \\
\end{longtable}
}

\end{document}